\documentclass[pra,twocolumn,showpacs,superscriptaddress]{revtex4}
\usepackage{graphicx,amsmath,amssymb,amsfonts,latexsym,color}
\usepackage[plainpages=false,hyperfootnotes=false,colorlinks=false]{hyperref}
\usepackage[normalem]{ulem}
\usepackage{dsfont}
\usepackage{array}

\newcommand{\vv}[1]{\mathbf{#1}} 

\newcommand{\ii}{\mathrm{i}}
\newcommand{\expo}[1]{\mathrm{e}^{#1}}
\newcommand{\pdfrac}[2]{\frac{\partial #1}{\partial #2}}
\newcommand{\T}[1]{\mathcal{T}\big[{#1}\big] }
\newcommand{\omegap}{\omega_\mathrm{p}}
\newcommand{\omegasp}{\omega_\mathrm{sp}}
\newcommand{\captionbf}[1]{\textnormal{\textbf{#1}}}
\mathchardef\mhyphen="2D

\begin{document}

\title{Plasmonic modes in nanowire dimers: A study based on 
       the hydrodynamic Drude model including nonlocal and nonlinear effects} 

     \author{Matthias Moeferdt}
\affiliation{Humboldt-Universit\"at zu Berlin, Institut f\"ur Physik, AG Theoretische Optik \& Photonik, 12489 Berlin, Germany}

     \author{Thomas Kiel}
\affiliation{Humboldt-Universit\"at zu Berlin, Institut f\"ur Physik, AG Theoretische Optik \& Photonik, 12489 Berlin, Germany}

     \author{Tobias Sproll}
\affiliation{Max-Born-Institut, 12489 Berlin, Germany}

 \author{Francesco Intravaia}
\affiliation{Max-Born-Institut, 12489 Berlin, Germany}

     \author{Kurt Busch}
\affiliation{Humboldt-Universit\"at zu Berlin, Institut f\"ur Physik, AG Theoretische Optik \& Photonik, 12489 Berlin, Germany}
\affiliation{Max-Born-Institut, 12489 Berlin, Germany}

\newcommand{\mathbfh}[1]{\hat{\mathbf{#1}}}

\begin{abstract} 
A combined analytical and numerical study of the modes in two distinct plasmonic nanowire 
systems is presented. The computations 
are based on a Discontinuous Galerkin Time-Domain approach and a fully nonlinear 
and nonlocal hydrodynamic Drude model for the metal is utilized. In the linear regime, 
these computations demonstrate the strong influence of nonlocality on the field distributions 
as well as on the scattering and absorption spectra. 
Based on these results, second-harmonic generation efficiencies are computed over a frequency 
range that covers all relevant modes of the linear spectra. In order to interprete the 
physical mechanisms that lead to corresponding field distributions, 
the associated linear quasi-electrostatic problem is solved analytically via conformal transformation 
techniques. This provides an intuitive classification of the linear excitations of the systems 
that is then applied to the full Maxwell case. Based on this 
classification, group theory facilitates the determination of the selection rules for the 
efficient excitation of modes, both in the linear and the nonlinear regime. This 
leads to significantly enhanced second-harmonic generation via judiciously exploiting
the system symmetries. 
These results regarding the mode structure and second-harmonic generation are of direct 
relevance to other nano-antenna systems.
\end{abstract}

\maketitle

\section{Introduction}
Nano-plasmonic structures lie at the heart of numerous recent advances in different research 
areas of fundamental physics and technological applications. Examples include surface- and tip-enhanced 
Raman scattering, frequency conversion, nano-antennas, metasurfaces, and hyperbolic metamaterials. While 
in many cases the modeling of the metallic elements via linear Drude- or Drude-Lorentz models is sufficient, 
there exist a number of scenarios where more refined material models are required. These scenarios include 
structures with nano-scale features such as nano-gaps, tips and grooves as well as nano-antennas with emitters 
in their respective near-field for which taking into account of the spatially nonlocal and/or 
nonlinear characteristics of the metals is necessary for an accurate description of the system. 
As a result, corresponding advanced materials models, specifically the hydrodynamic Drude model
and various extensions thereof, have recently received considerable attention
\cite{sipe1980surface,liebsch1993surface,halevi1995hydrodynamic,garcia2008nonlocal,villoperez2009hydrodynamical,
      ciraci2012origin,ciraci2012second,stella2013performance,ciraci2013hydrodynamic,
      mortensen2014generalized,christensen2014nonlocal,raza2015nonlocal,toscano2015resonance,
ciraci2016quantum,scalora2010second,zeng2009classical,hille2016second,huynh2016ultrafast,toscano2012surface,toscano2012modified}.
For instance, the hydrodynamic Drude model allows to investigate the frequency shifts, the
excitation of bulk plasmons, nonlinear wave-mixing phenomena such as second harmonic generation,
and modified field distributions and intensities. While it has been argued that, in the linear 
regime, the hydrodynamic model yields only semi-quantitative results~\cite{stella2013performance},
recent improvements regarding refined treatments of the systems' behavior at interfaces~\cite{toscano2015resonance,ciraci2016quantum},
as well as the incorporation of interband transitions~\cite{liebsch1993surface,toscano2015resonance}
and Landau damping~\cite{halevi1995hydrodynamic,mortensen2014generalized} have demonstrated that
fully quantitative results may be obtained nonetheless~\cite{toscano2015resonance}. 

Numerical computations using the hydrodynamic Drude model are more demanding than those
employing the ordinary Drude model. In particular, the description of the interfaces' 
behavior as well as of the potential constraining the electronic fluid, needed for an 
adequate resolution of bulk plasmons with typical wavelength of the order of 1 nm,
requires very efficient Maxwell solvers. In recent years, new numerical schemes for solving the hydrodynamic equations in 
a non-perturbative fashion have emerged \cite{ginzburg2014nonperturbative,hille2016second}.
In contrast to earlier works~\cite{ciraci2012origin,ciraci2012second},
time-domain approaches provide direct access to the nonlinear properties 
of the hydrodynamic Drude model by solving the full set of nonlinear equations~\cite{hille2016second,huynh2016ultrafast,ginzburg2014nonperturbative}, 
without making any further assumptions about the nonlinear source terms.
In particular, the Discontinuous Galerkin Time-Domain 
(DGTD) method~\cite{busch2011discontinuous}, a time-dependent finite-element framework, has shown 
good performance characteristics both for nonlocal and nonlinear properties~\cite{hille2016second,schmitt2016dgtd}.

In this work, we utilize the DGTD approach to investigate the hydrodynamic Drude model. Specifically, we discuss 
both its nonlocal and nonlinear characteristics by employing a cylindrical nanowire dimer system
and a single V-groove structure as illustrative examples. 
The cylindrical dimer is a particularly interesting system to study: 
Indeed, despite its low symmetry, it can be treated analytically within a (quasi-)electrostatic approach. 
This allows, in turn, to employ 
a conformal transformation which provides a quite useful classification of the modes existing in the 
system (see Section \ref{sec:electrostatic}). These analytical results are subsequently utilized to 
accurately interpret the numerical outcomes (Section \ref{NumericalResults}). 
Our first key finding 
is, that for a specific angle of incidence, the spatial nonlocality has a particularly strong effect 
on the scattering spectra. Secondly, we demonstrate that, due to the nonlinearity, modes are efficiently 
excited via second harmonic generation (SHG), while in the linear regime their excitation is symmetry-suppressed. 
In addition, our group-theoretical considerations allow to discuss SHG efficiencies by properly 
tuning the modes into singly- and doubly-resonant conditions. These key findings are directly 
applicable to more general nano-antenna structures.

\section{Quasi-electrostatic Theory of a Cylindrical Nanowire Dimer}
\label{sec:electrostatic}
\begin{figure}
  \includegraphics[width=\linewidth]{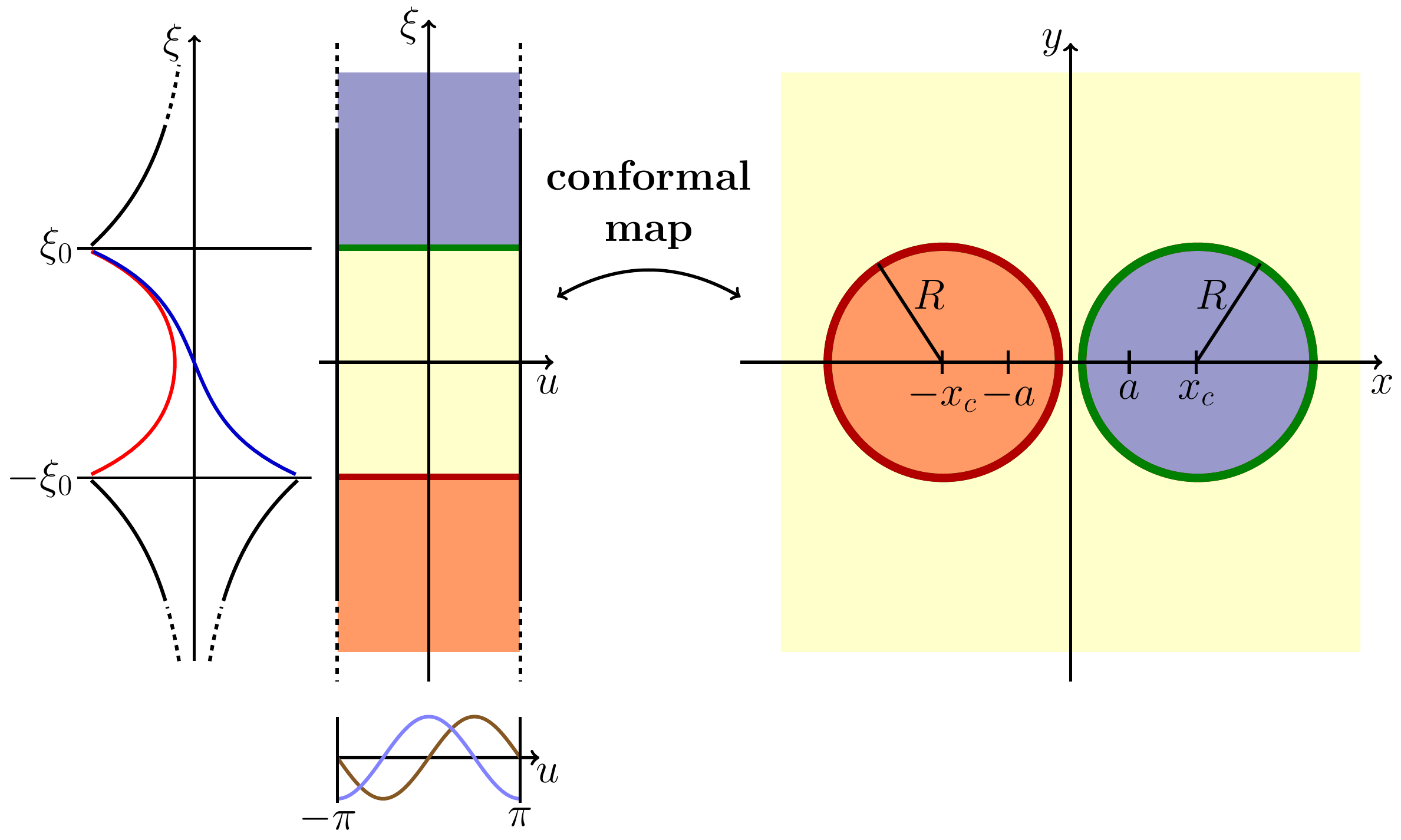} 
  \caption{Conformal map for bicylindrical coordinates. The colors indicate which regions are mapped
           onto each other. The cylindrical structure imposes a $2\pi$-periodicity on the $u$-coordinate. 
	 		 		 The solutions for the electrostatic potential, as functions of the bicylindrical coordinates 
					 $\xi$ and $u$, are therefore given by trigonometric and hyperbolic functions (displayed on 
					 the left).
					 \label{fig:conformal_map}}
\end{figure}

We consider two identical, parallel, and infinitely extended circular nano-wires of 10 nm radius, 
separated by a 2 nm gap. These wires are situated in vacuum and are made of a Drude metal,
described by a (local and linear) dielectric constant 
\begin{align}
 \epsilon(\omega) = 1 - \frac{\omegap^2}{\omega (\omega + \ii \gamma)}, 
 \label{eq:drudelocal}
\end{align}
with a plasma frequency $\omega_p=1.39\cdot10^{16}$ rad/s ($\sim 9$ eV) and damping constant 
$\gamma=3.23\cdot10^{13}$ rad/s ($\sim 21$ meV).
These values correspond to the Drude-contribution to the permittivity of silver reported by Johnson and Christy~\cite{johnson1972optical}.
With these parameters the system can be described to a good approximation within the 
quasi-electrostatic approximation, which amounts to neglecting retardation effects and 
setting the speed of light to infinity.
In the quasi-electrostatic limit, the dimer structure can be treated analytically. This is 
done by performing a conformal map to an appropriate coordinate system, where the Laplace 
equation, which in quasi-electrostatics takes over the role of the wave equation,  
becomes separable~\cite{rimrock,ruppin1989optical,luo2013surface}. The mapping, which describes a 
transformation between a strip of $\mathbb{R}^2$ and two cylinders, is depicted schematically 
in Fig.~\ref{fig:conformal_map}. 
The corresponding bicylindrical coordinates are called $\xi$ and $u$ and the transformation 
from the cartesian coordinates reads
\begin{align}
x = \frac{ a \sinh{\xi}}{\cosh{\xi} - \cos{u}}, \quad y = \frac{ a \sin{u}}{\cosh{\xi} - \cos{u}}.
\end{align}
The system is built around two foci which lie at $x=\pm a$. The parameter $a$ depends 
on the parameters characterizing the system we wish to describe and is uniquely determined 
for a specific set of radii, $R_{1,2}$, and center points, $x_{c_{1,2}}$, of the cylinders
by the following relations~\cite{rimrock}
\begin{align}
 x_{c_{1,2}}=\frac{a}{\tanh{\xi_{1,2}}} ,\quad  R_{1,2}=\frac{a}{|\sinh{\xi_{1,2}}|} 
 \label{rundx}.
\end{align}
The $\xi$-level lines are non-concentric circles (so-called 
Apollonian circles) around the foci, which lie in the left half space ($x<0$) for $\xi<0$ and 
in the right half space ($x>0$) for $\xi>0$. For $\xi  \rightarrow \pm \infty$ the circles 
collapse into the foci and for $\xi \rightarrow 0$ they enclose the corresponding half 
space ($x>0$ for $\xi \rightarrow 0^+$ or $x<0$ for $\xi \rightarrow 0^-$) and the entire 
$y$-axis. 
The vacuum-metal interfaces where the boundary conditions for the fields have to be imposed 
are thus given by two fixed $\xi$-level lines, $\xi=\xi_{1,2}$. In the present case, we consider 
cylinders with equal radii $R_{1,2} = R=10$ nm and a gap of $2$ nm, corresponding to 
$x_{c_{1,2}}=\pm x_c =\pm11$ nm. We then have
\begin{align} \xi_{1,2}&=\pm \xi_0 = \pm \mathrm{arcosh}\left(\frac{x_c}{R}\right)=\pm0.443568245... \\
 a&=4.582575... \mathrm{nm}.
\end{align}
Within these coordinates, the Laplace equation reads
\begin{align}
 \Delta V = \left(\frac{\cosh{\xi}-\cos u}{a}\right)^2\left[\partial_\xi^2 V + \partial_u^2 V \right] = 0.
\end{align}
This equation is separable and has the harmonic solutions
\begin{align}
  V_1 &= \left[\sinh{m\xi},\cosh{m\xi}\right]\hspace{.3cm} 
	       \mathrm{or} \hspace{.3cm}
				 \left[\expo{m\xi},\expo{-m\xi}\right], \label{eq:xiharmonics} \\
  V_2 &= \left[\sin{mu},\cos{mu}\right]\hspace{.62cm} 
	       \mathrm{or} \hspace{.3cm}
				 \left[\expo{\ii m u},\expo{-\ii m u}\right], \label{eq:uharmonics}
\end{align}
which are depicted along with the mapping in Fig.~\ref{fig:conformal_map}. 
From this, we can construct four physical solutions as combinations of $u$ and $\xi$, where
the variable $u$ plays the role of the azimuthal coordinate and introduces a $2\pi$-periodicity 
which results in a ``quantization'' of the solutions labeled by the discrete index $m \in \mathbb{N}^+$. 

In the quasi-electrostatic description the electromagnetic boundary conditions read
\begin{align}
                                                  V_i(\pm\xi_0) &= V_o(\pm\xi_0), \notag \\
  \left.\epsilon(\omega)\pdfrac{V_i}{\xi}\right\vert_{\pm\xi_0} &= \left.\pdfrac{V_o}{\xi}\right\vert_{\pm\xi_0}, 
	\label{eq:boundaryconditions}
\end{align}
where the subscripts $i$ and $o$
stands for inside and outside of the cylinder, respectively. The permittivity on the inside is 
given by Eq.~\eqref{eq:drudelocal}. 
The hyperbolic functions of Eqs.~\eqref{eq:xiharmonics} are chosen to describe the field outside the 
cylinders, while exponentially decaying solutions characterize the field inside the cylinder 
(see Fig.~\ref{fig:conformal_map}).
\begin{figure}[ht]
	\includegraphics[width=\linewidth]{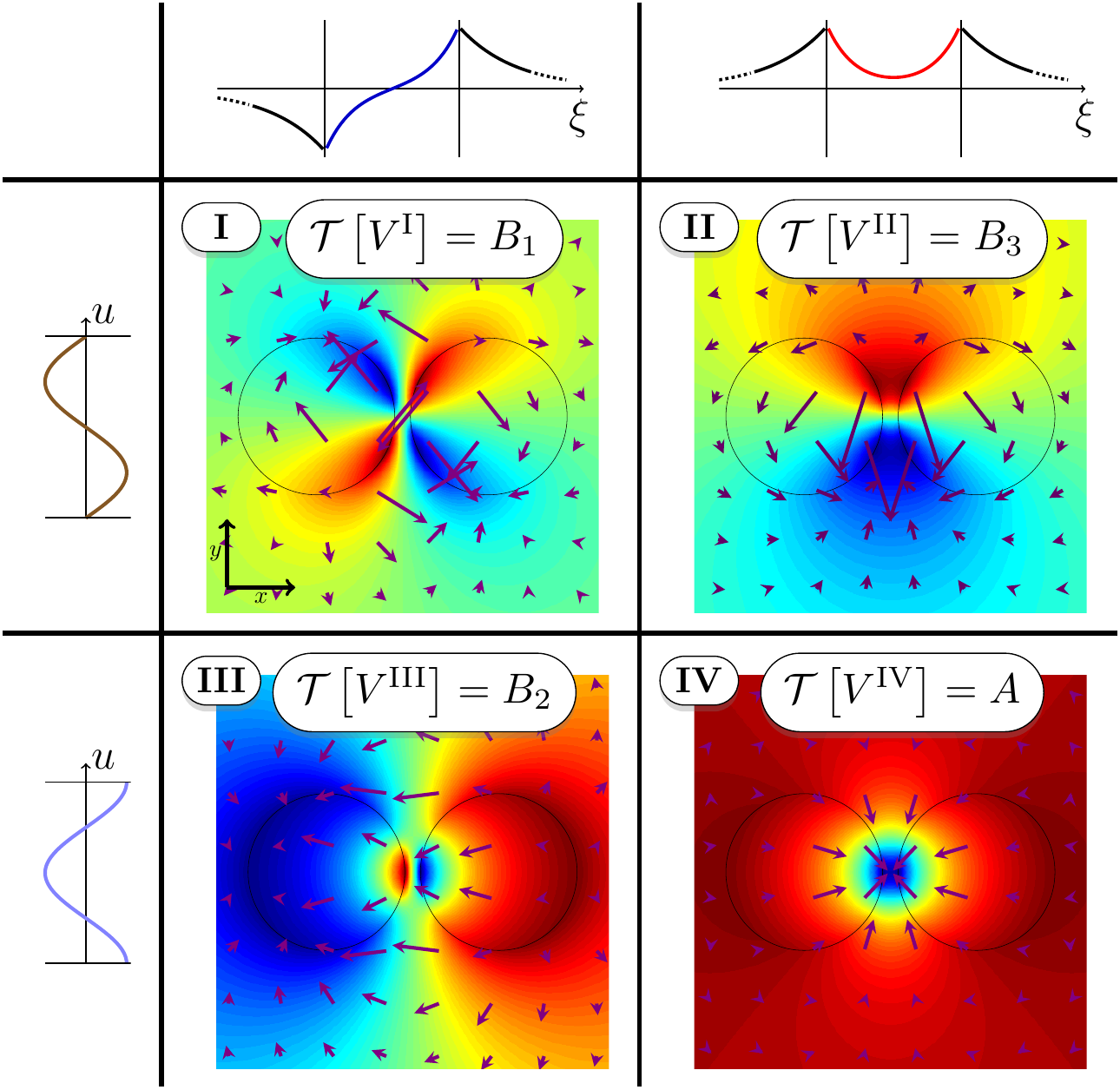}
	\caption{The harmonic solutions for the electrostatic potential in a bicylindrical
					 setup allow for four classes of solutions with different symmetries. The 
					 potentials for $m=1$ are displayed for each solution, the arrows depict the 
					 gradients of the potentials. 
					 For each potential, the irreducible representation is given.
					 	\label{fig:classes_of_potentials}}
\end{figure}
As a consequence, the composite solutions exhibit different symmetry properties and we categorize 
them in four classes (labeled I-IV). The solutions for the potentials and those for the resulting 
electric field distributions are displayed in Figs.~\ref{fig:classes_of_potentials} 
and~\ref{fig:classes_of_solutions}.
\begin{figure}[ht]
 \includegraphics[width=\linewidth]{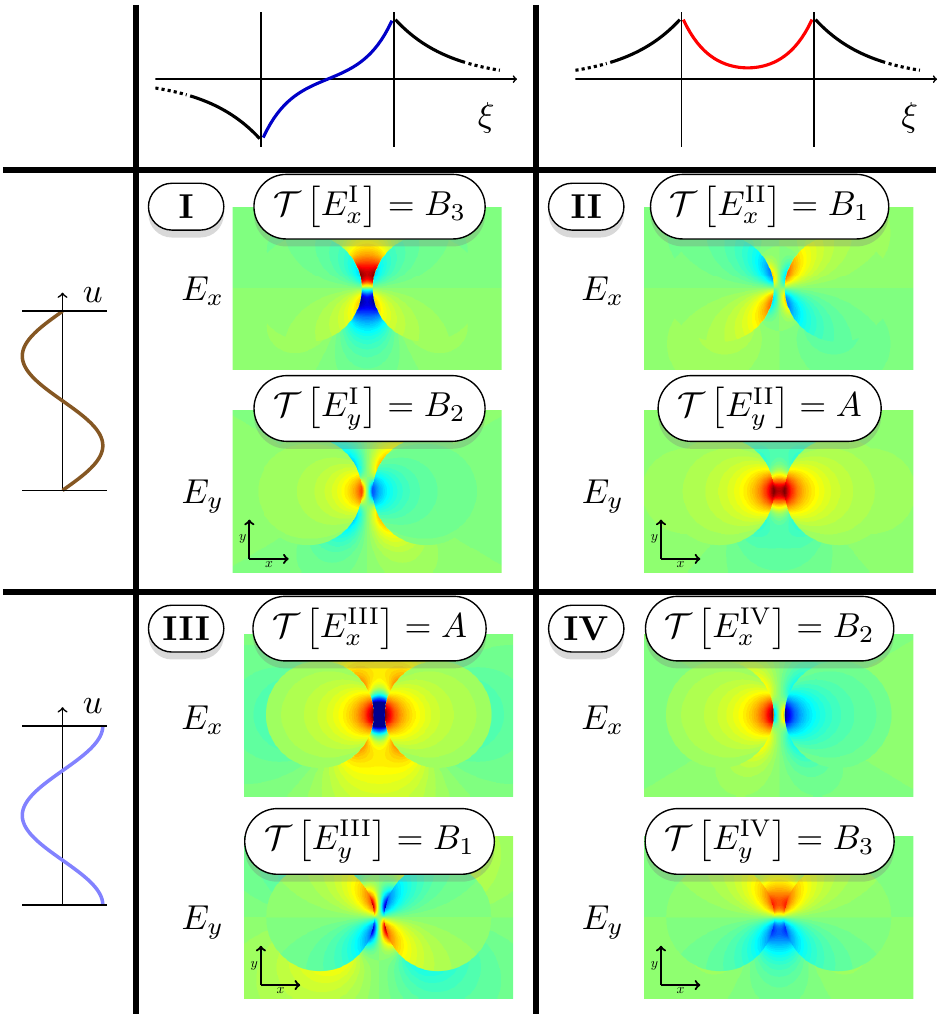}
	\caption{Electric field distributions for the $m=1$ modes pertaining to the four 
           symmetry classes. For each field component, the irreducible representation 
					 is given. The field components were calculated directly from the potentials 
           in Fig.~\ref{fig:classes_of_potentials} via $\vv{E}=-\nabla V$.
           	\label{fig:classes_of_solutions}}
\end{figure}
The four classes of potentials from Fig.~\ref{fig:classes_of_potentials} can be identified 
with the four irreducible representations of the dihedral symmetry point group $D_{2}$ (in 
two dimensions, $D_{2}$ is isomorphic to the group $C_{2v}$)~\cite{gericke2016charakteren,SuppMat}. 
There are four symmetry operations in this group, all of which map the original dimer 
geometry onto itself: The identity operation ($I$), a rotation around the $z$-axis by 180 
degrees ($C_{2z}$), mirroring on the $xz$-plane ($\sigma_{xz}$) and mirroring on the $yz$-plane
($\sigma_{xz}$).
The group is Abelian and the character table of $D_2$ reads~\cite{gericke2016charakteren}:
 \begin{center}
  \begin{tabular}{|>{\centering}m{1.5cm} | >{\centering}m{1.5cm}| >{\centering}m{1.5cm}| >{\centering}m{1.5cm} | > {\centering\arraybackslash}m{1.5cm}|}
		\hline $ D_2$ & $I$ & $C_2(z)$ & $\sigma_{xz}$ & $\sigma_{yz}$ \\ \hline 
			$A	$ & $1$ &  $1$ &  $1$ &  $1$ \\ \hline
			$B_1$ & $1$ &  $1$ & $-1$ & $-1$ \\ \hline 
			$B_2$ & $1$ & $-1$ &  $1$ & $-1$ \\ \hline
			$B_3$ & $1$ & $-1$ & $-1$ &  $1$ \\ \hline
	\end{tabular}
\end{center}
Consequently, the four potentials transform according to
\begin{align}
  \T{V^{\mathrm{I}}}   = B_1, \quad & \T{V^{\mathrm{II}}} = B_3, \notag \\
  \T{V^{\mathrm{III}}} = B_2, \quad & \T{V^{\mathrm{IV}}} = A,
\end{align}
where $\T{F}$ denotes the decomposition of a quantity $F$ into irreducible representations, 
i.e., it indicates under which transformation $F$ is mapped onto itself. The two-dimensional 
Nabla-operator transforms as
\begin{align}
	\T{\nabla} = \begin{pmatrix} \T{\partial_x} \\ \phantom{.} \\ 
	                             \T{\partial_y} 
							 \end{pmatrix} 
						 = \begin{pmatrix} B_2 \\ 
						                   B_3 
							 \end{pmatrix}.
\end{align}
The multiplication rules can be readily extracted from the character table by multiplying 
the elements in the same column and identifying the results with one of the four possible 
symmetry classes. By applying the Nabla-operator to the potentials
\begin{align}
 \T{\nabla}\T{V^i} = \T{\vv{E}^i} \, \, ,\quad i=\mathrm{I}...\mathrm{IV}, 
\end{align}
we obtain the symmetries of the fields in Fig.~\ref{fig:classes_of_solutions}, which are 
not equivalent to the symmetries of the potentials from the same class:
\begin{align}
	\T{\vv{E}^\mathrm{I}} = \begin{pmatrix} B_3 \\ 
	B_2  \end{pmatrix}, 
	\quad &
	\T{\vv{E}^\mathrm{II}}  = \begin{pmatrix} B_1 \\ 
	A  \end{pmatrix}, \notag \\
	\T{\vv{E}^\mathrm{III}} = \begin{pmatrix} A \\ 
	B_1  \end{pmatrix},  
	\quad &
	\T{\vv{E}^\mathrm{IV}} = \begin{pmatrix} B_2 \\ 
	B_3  \end{pmatrix}. 
	\label{feldsymmetrien}
\end{align}
In order to excite one of the modes, the irreducible representation of the electric fields 
of an incoming light pulse has to match those of the field components in Eq.~\eqref{feldsymmetrien}. 
A static external field (as one would have it within a capacitor) is represented by $A$. 
In consequence, a static electric field in $x$-direction is capable of exciting the mode 
from class II and a static field in $y$-direction that of class III. 

By inspection, one finds that the electric field $\vv{E}^{k_x}$ of a wave-packet propagating 
in $x$-direction, with an electric field in $y$-direction, is represented by
\begin{align}
	\T{E^{k_x}} = A \oplus B_2 ,
	\label{excitation1}
\end{align}
where the direct-sum-symbol $\oplus$ indicates that the symmetries of the field are given 
by a superposition of two irreducible representations. In this sum, all signs and magnitudes 
are dropped. Note however, that the deeper we enter the quasi-electrostatic limit (smaller 
particle sizes), the smaller the influence of the $B_2$-contribution on the spectra 
becomes. The type of excitation given by Eq.~\eqref{excitation1} complies with the 
symmetries of the modes II and I.

On the other hand, for a pulse propagating in $y$-direction with an electric field in 
$x$-direction, we have an electric field
\begin{align}
	\T{E^{k_y}} = A \oplus B_3.
\end{align}
This field, therefore, excites modes of class I and III (cf. Fig.~\ref{fig:classes_of_solutions}). 
Thus, modes of class IV are dark for wave packages of this type or linear combinations thereof, 
i.e. regardless of the angle of incidence. 

\begin{figure}[ht]
	\includegraphics[width=\linewidth]{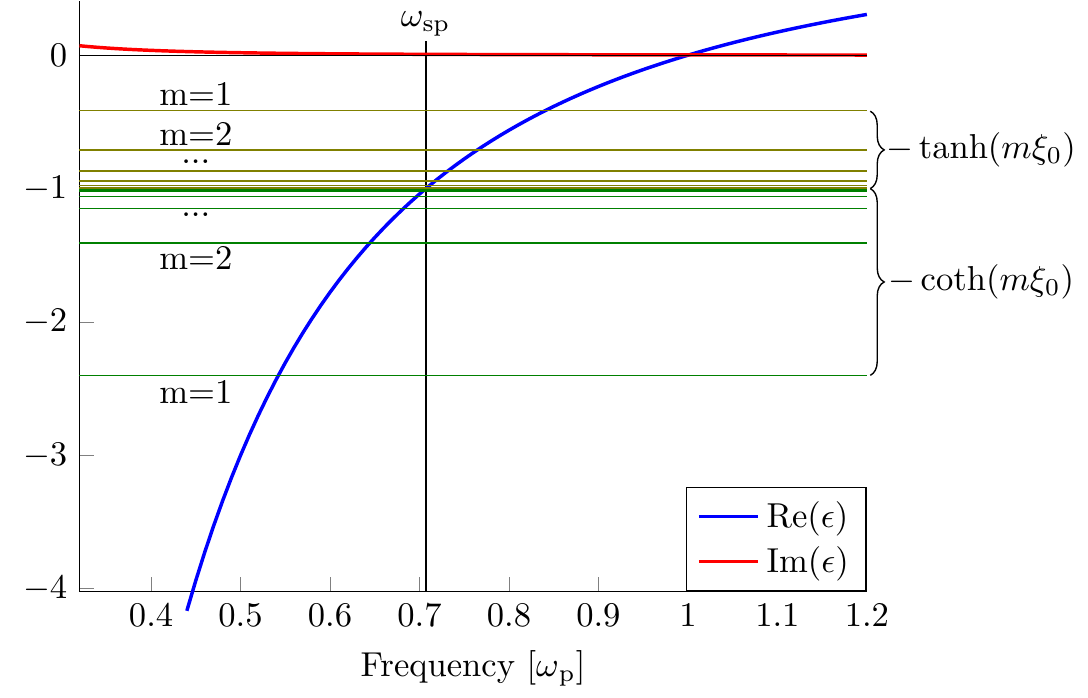}
	\caption{Graphic solution for the electrostatic problem for the cylindrical dimer (neglecting 
	         the imaginary part of the permittivity). We find a hybridization of the solutions. 
					 For large $m$, the modes tend to the surface plasmon frequency $\omega_{\mathrm{sp}}$.
					 \label{fig:graphic_solution}}
\end{figure}

Having discussed the symmetries of the modes, we proceed to the determination of their frequencies. 
In the quasi-electrostatic problem, the frequency-dependence is introduced by means of the material 
model and the boundary conditions given by Eq.~\eqref{eq:boundaryconditions}. 
In these considerations, the azimuthal variable $u$ does not play a role, because the angular 
dependence is the same everywhere in space. Therefore, within the quasi-electrostatic theory, 
classes I and III as well as classes II and IV are degenerate. As a result, we arrive at the 
following solutions:
\begin{align}
	\epsilon(\omega) &= - \coth(m \xi_0) \quad \, \mathrm{for} \,\mathrm{solutions}\,\, \mathrm{I,III}, \notag \\
	\epsilon(\omega) &= - \tanh(m \xi_0) \quad \, \mathrm{for} \,\mathrm{solutions}\,\, \mathrm{II,IV}, 
	\label{eq:electrostatic_modes}
\end{align}
which are displayed in Fig.~\ref{fig:graphic_solution}. (The imaginary part of the permittivity
is neglected in the graphical solution, but it is included in the analytical calculation of the 
mode frequencies). 
The degeneracy is lifted when we go beyond the quasi-electrostatic limit.
As demonstrated in Fig.~\ref{fig:graphic_solution}, for a cylindrical dimer we find a hybridization 
of modes, having frequencies both below and above that of a single cylinder, namely the surface 
plasmon frequency $\omega_{\mathrm{sp}}=\omegap/\sqrt{2}$, which is obtained from 
$\mathrm{Re}(\epsilon)=-1$. 
The modes for $m=1$ are well separated from the next higher modes. For large $m$, the mode frequencies 
tend to $\omega_{\mathrm{sp}}$. For infinite separation, 
$x_c \rightarrow \infty$ the mode frequencies collapse to $\omegasp$ and the field distributions 
become the well-known multipolar distributions of regular cylinders, with their dipoles oriented 
according to the symmetries described above.

\section{Numerical Results}
\label{NumericalResults}
\subsection{CylindricalDimer}

The hydrodynamic model describes the electrons in the metal as a charged and compressible 
fluid subject to electromagnetic forcing, leading to a set of equations of motion 
comprising the Euler equation and the equation of charge conservation. The full electromagnetic 
problem is then defined by the solution of the dynamics of this system along with the 
Maxwell equations that couple to the fluid equations via the electric current.
We employ the usual hard-wall slip boundary condition, which consists in setting the normal 
component of the fluid's velocity to zero at the material surface, allowing the electrons 
to slip sideways \cite{Hesthaven_2007}. 
This additional boundary condition (of non-electromagnetic 
origin), which is equivalent to the vanishing of the normal component of the
current at the interface, is physically related to the possibility to have a nonzero volume 
charge-density at the surface
~\cite{jewsbury1981electrodynamic,ford1984electromagnetic}. The pressure term appearing 
in the hydrodynamic equation (see below) induces a smearing out of the charge-density on 
the scale of the plasma screening length (given in our case by the Thomas-Fermi wavelength)
~\cite{manfredi2005model}, preventing the fluid to have singular behaviors at the surface.
The influence of the resulting electron spill-out at the particle surface has previously 
been studied by Toscano \textit{et al.} in Ref. \cite{toscano2015resonance} 
in a framework similar to the one employed here.

The nonlinear, nonlocal and fully retarded version of the hydrodynamic model is described by the following 
equations~\cite{hille2016second}
\begin{subequations}
\begin{align}
       \partial_t n  & = - \nabla \cdot (n\vv{v}),  \label{continuity} \\
 n \partial_t \vv{v} & = - n (\vv{v} \cdot \nabla) \vv{v} - \frac{1}{M} \nabla p   
                         - \gamma n \vv{v} +  \notag \\
                     & \phantom{=} 
 + \frac{e}{M} n \left(\vv{E} + \mu_0 \vv{v} \times \vv{H} \right),
  \label{momentum}
\end{align}
\label{HydroEqs}
\end{subequations}
\noindent
where $M=0.96\,m_\mathrm{e}$ ($m_\mathrm{e}$ is the free electron mass~\cite{johnson1972optical}),  
$n=n(\mathbf{r},t)$, $\vv{v}=\vv{v}(\mathbf{r},t)$ and $p=p(\mathbf{r},t)$ denote the electron 
fluid density, velocity and pressure, respectively.
Equations \eqref{HydroEqs} are conservation relations addressing the number of charge carriers 
[Eq. \eqref{continuity}] and the their total momentum [Eq. \eqref{momentum}]. They have been 
applied in different circumstances to describe the dynamics of carriers not only in metals 
but also semiconductors~\cite{sumi1967traveling,thiennot1972ondes}.
For metals, using the Thomas-Fermi approximation, the 
pressure is given as
\begin{align}
 p = \frac{1}{5}\frac{\hbar^2}{m_e}(3\pi^2)^{2/3} n^{5/3}.
\end{align}
Upon linearization, the above set of equations for the longitudinal component give the 
usual nonlocal permittivity
\begin{align}
  \epsilon_{\rm L}(\vv{k},\omega) = 1-\frac{\omegap^2}{\omega(\omega+\ii\omega\gamma) -\beta^2 k^2},  
	\label{nonlocal_permittivity}
\end{align}
while the transverse permittivity, $\epsilon_{\rm T}(\omega)$, has still the local form 
of Eq. \eqref{eq:drudelocal}.
For our set of parameters we have $\beta^2\approx 6.9738\cdot 10^{11} {\rm m}^{2}/{\rm s^{2}}$.
Owing to its simplicity, the linearized model allows for an analytical solution of the full 
electromagnetic problem in certain simple geometries. This is the case for a single-cylinder 
configuration, which has been used as an analytical reference for the convergence study of 
our numerical computations~\cite{hille2016second}.
In the following, the DGTD method is used to solve the full nonlinear and nonlocal electromagnetic 
problem and to compute the scattering spectra of the dimer system.
To calculate the spectra, a total-field/scattered-field (TFSF) formalism is employed, wherein 
a closed contour is provided around the scatterer. At this contour a pulse is injected and 
the scattering signal is subsequently calculated by recording the flux of the Poynting through 
said contour~\cite{Hesthaven_2007,busch2011discontinuous}.
At first,
the system is excited using a spectrally broad-band Gaussian pulse with central frequency 
$0.67\,\omega_\mathrm{p}$, a width of $1.36\,\omegap$ and a rather low field amplitude of 
$10^3\,$V/m. The $\vv{E}$-field is polarized in-plane and we study two angles of 
incidence -- incidence perpendicular to the dimer axis ($\vv{k}$ in $y$-direction) and along 
the dimer axis ($\vv{k}$ in $x$-direction).
\begin{figure}[!t]
	\includegraphics[width=\linewidth]{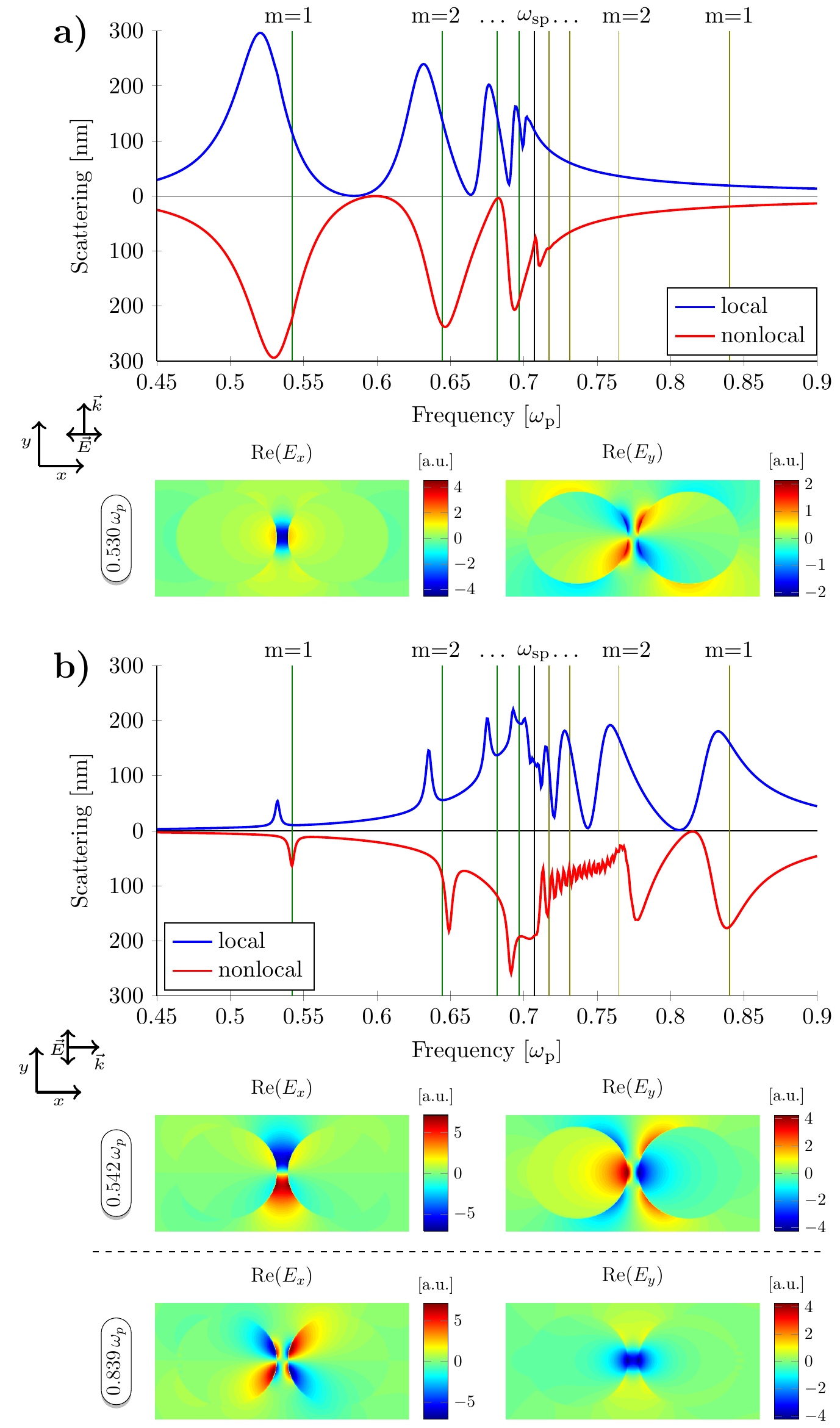}
	\caption{The local and nonlocal spectra for a dimer structure for 
	         \textbf{a)} incidence perpendicular to the dimer axis 
					 \textbf{b)} incidence along the dimer axis. 
					 The nonlocality has a particularly strong effect for case \textbf{b)}. 
					 For the local Drude model, the spectrum in \textbf{a)} contains
           exclusively modes of class I or III that are confined to frequencies 
					 below the surface plasmon frequency $\omega_\mathrm{sp}$ (as expected 
					 from (quasi-)electrostatic theory). 
					 For the nonlocal hydrodynamic Drude model, the entire spectrum is 
					 blue-shifted. The field distributions were recorded at the frequencies 
	         corresponding to the $m=1$ resonances and can be identified with the 
					 (quasi-)electrostatic solutions from Fig.~\ref{fig:classes_of_solutions}. 
					 The positions of the quasi-electrostatic modes, according to 
					 Eq. \eqref{eq:electrostatic_modes}, are indicated by vertical lines. 
					 Only the real part of the field distributions is displayed. The 
					 imaginary part can be found in the Supplemental Material \cite{SuppMat}.
					 \label{fig:spectra}}
\end{figure}
In Fig.~\ref{fig:spectra},
we display the
scattering spectra, calculated using a
nonlocal hydrodynamic Drude
model on the one hand, and a
local Drude
model on the other hand, while indicating the
analytic quasi-electrostatic solutions
by vertical lines \cite{Note1}. 
The corresponding spectra
for both the local and the nonlocal calculation
exhibit clearly separated peaks which allow for a classification by identifying them with 
the quasi-electrostatic solutions for the different symmetry classes and different values 
of $m$ (see Sec. \ref{sec:electrostatic}).
The nonlocal spectra exhibit a blueshift with respect to the local calculations, which is 
a typical feature of the nonlocal material model in conjunction with hard-wall boundary 
conditions
~\cite{ruppin2001extinction}.

In Fig.~\ref{fig:spectra} \textbf{a)}, we display the spectra for an incident pulse 
propagating along the dimer axis. As expected from our group-theoretical analysis, 
for incidence along the dimer axis, all modes should be of either class I or class III 
and, therefore, lie below the plasmon frequency.
In fact, the fields belonging to the prominent peaks can be identified as belonging 
to class III. Modes of class I are, however, also allowed.
Within the quasi-electrostatic description, they are degenerate with those of 
class III and are, therefore, not visible in the scattering spectrum. In the 
Supplemental Material \cite{SuppMat}, we show that the modes of class I are, in fact, also 
excited and that in a fully retarded computation the degeneracy in frequency is 
lifted.

For incidence along the dimer axis (Fig.~\ref{fig:spectra} \textbf{b)}), we have 
solutions of classes I and II,
which, respectively, are energetically lying below and above the surface plasmon frequency.
This also confirms the expectations from our group-theoretical considerations above
\cite{Note2}. 
For incidence along the dimer axis, we observe another notable effect of the 
nonlocality: The spectra show a number of small but well-separated peaks just 
above the surface plasmon frequency $\omega_{\mathrm{sp}}$ which do not exist 
in the local case. They originate from high-order modes which, in a local 
description, coalesce at the limiting point $\omegasp$ while, within a 
nonlocal framework, split away and are increasingly blue-shifted beyond 
this frequency, separating from each other
~\cite{christensen2014nonlocal}.

\begin{figure}[ht]
	\includegraphics[width=\linewidth]{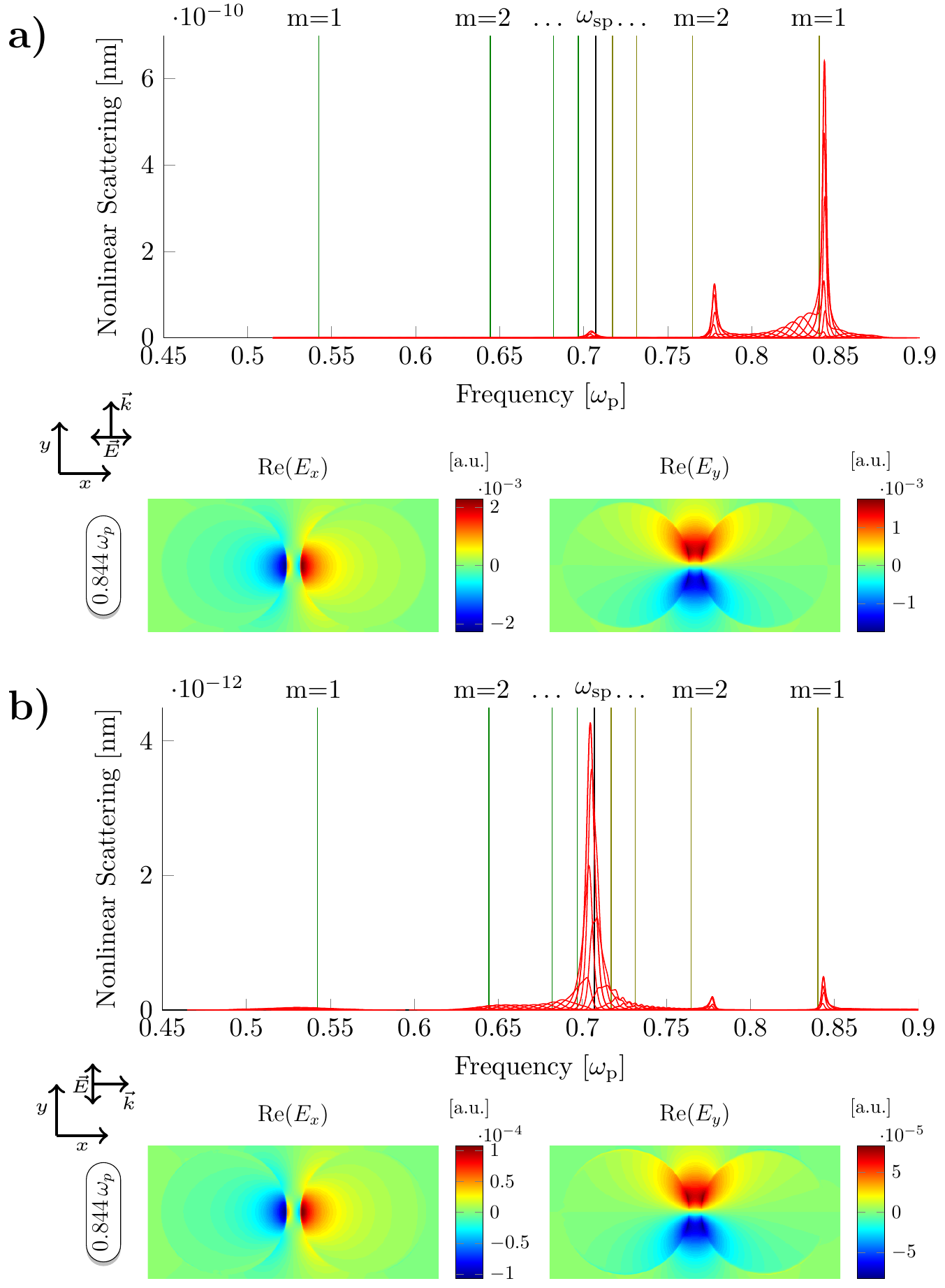}
	\caption{Second harmonic signals for 
	         \textbf{a)} incidence perpendicular
           and for \textbf{b)} incidence along the dimer axis, 
					 obtained via a scan using a sequence of spectrally narrow-band pulses at  the
					 fundamental frequencies. The field distributions are displayed for the $m=1$ mode. 
					 The vertical lines indicate the positions of the quasi-electrostatic modes, 
					 see Eq.~\eqref{eq:electrostatic_modes}.}.
					 \label{fig:shg_spectra}
\end{figure}

As a next step, we investigate the nonlinear properties of our model. As it 
was done in Ref. \cite{hille2016second}, we raster scan a broad frequency range, 
by exciting the system with spectrally narrow-band Gaussian pulses (each with a 
spectral width of $0.03\,\omegap$ and a  field strength of $10^6\,$V/m)
that are centered at
different fundamental frequencies and subsequently recording the second harmonic 
signal once again employing the TFSF-formalism. 
Specifically, the excitation pulses at the fundamental frequencies have to be 
chosen to exhibit a sufficiently narrow bandwidth such that their spectra are
have zero overlap with second harmonic frequency range, thereby enabling a
background-free detection of the second harmonic signal.
In Fig.~\ref{fig:shg_spectra}, we display the resulting second harmonic 
spectra (The signal at the fundamental frequency of the excitation is {\it not} displayed 
in the figure, it matches the linear spectrum known from the broad-band 
calculations above).
The corresponding SHG peaks can once again be associated with certain well-defined 
$m$-values of the quasi-electrostatic theory. For both excitation directions, 
we find a strong signal near $0.85 \, \omega_\mathrm{p}$, which corresponds 
to an $m=1$ mode. Through the field distributions, we determine that the excitation
-- for both angles of incidence -- belongs to the symmetry class IV, which cannot be excited on the linear level 
from the far-field. Furthermore, for an excitation propagating
perpendicular to the dimer axis (Fig.~\ref{fig:shg_spectra} \textbf{a)}), only resonances 
above the surface plasmon frequency $\omegasp$ occur. There is a broad resonance 
just below the class IV $m=1$ mode. This broad peak corresponds the class II $m=1$ 
mode.

For an incident pulse along the dimer axis  (Fig.~\ref{fig:shg_spectra} \textbf{b)}), there are resonances 
below and above the surface plasmon frequency $\omegasp$. An inspection of the field distributions 
shows that the belong to class III. 
This means that for both angles of incidence, we are now exciting those modes that were previously 
not excited in the linear calculations. 

A group-theoretical investigation of the intrinsic symmetries of the hydrodynamic Drude 
model in combination with the Maxwell equations provides an explanation for this behavior. 
In order to facilitate our analysis, we treat the combined nonlinear set of equations 
within a perturbative approach. Specifically, the fields pertaining to the hydrodynamic 
equations are expanded into a series of harmonics~\cite{sipe1980surface,huynh2016ultrafast}
\begin{align}
 \vv{E} &=  \vv{E}_0 + \vv{E}_1 + \vv{E}_1 + ...\\
 \vv{H} &=             \vv{H}_1 + \vv{H}_2 + ... \\
      n &=       n_0 +   n_1    +    n_2   + ...\\
 \vv{v} &=             \vv{v}_1 + \vv{v}_2 + ... ,
\end{align}
where the subscripts $i=0, 1, 2, ...$ correspond to static fields, fields at the
fundamental frequencies, and fields pertaining to the second-order response, respectively. 
Upon inserting the above expansions into the hydrodynamic equations~\eqref{HydroEqs}, 
we obtain source distributions for the fields at the second-order response that are
expressed in terms of various combinations of the fields at the fundamental frequency. 
Consequently, we can determine the symmetry properties of the second-order fields from 
the symmetries of the corresponding source distributions, i.e., from the symmetries of 
the fundamental fields through appropriate compatibility relations (see the Supplemental Material \cite{SuppMat}).
Using this procedure, we find
\begin{align}
 \T{E_{2x}} =  B_2 \left[\T{E_{1x}}\right]^2.  \label{eq:secondharmonicsymmetries}
\end{align}
We recall that for a wave-packet propagating in $y$-direction, the symmetry of the incoming 
electric field is given by 
\begin{align}
 \T{E^{k_y}_{x,\mathrm{inc}}} =  \T{E_{1x}}= A \oplus B_3,
\end{align}
which is compatible with modes of class I and II that are both excited in this case.
Calculating the second-order fields 
\begin{align}
  \T{E^{k_y}_{2x}} = B_2\left[A \oplus B_3 \right]^2 = B_2 (A \oplus B_3)  = B_2 \oplus B_1
\end{align}
reveals that
the second harmonic modes
are associated with the remaining two symmetries, 
i.e., the modes of class II and IV (cf. Fig.~\ref{fig:classes_of_solutions}). We thus explicitly see
 that the transformation behavior of the second-order fields differs from that
of the fundamental (excitation) fields, 
thus
explaining
many non-trivial features of SHG 
spectra.
The considerations on the incidence of a pulse propagating in the $x$-direction proceed 
in a completely analogous fashion. Starting with
\begin{align}
  \T{E^{k_x}_{y,\mathrm{inc}}} = A \oplus B_2,
\end{align}
the modes of class I and II are excited, which must have first order $E_x$-fields of the 
type
\begin{align}
  \T{E^{k_x}_{1x}} = B_1 \oplus B_3.
\end{align}
Therefore, for the second-order response, we find
 \begin{align}
 \T{E^{k_x}_{2x}} = B_2\left[B_1 \oplus B_3 \right]^2 = B_2 (A \oplus B_2)  = B_2 \oplus A,
\end{align}
which means that
second harmonics
of class III and IV are excited (cf. Fig.~\ref{fig:classes_of_solutions}).

\subsection{V-Groove}
\begin{figure}[ht]
	\centering
	\includegraphics[width=0.8\linewidth]{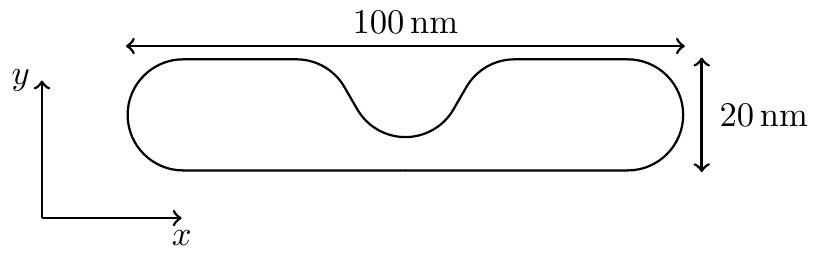}
	\caption{Schematic of the V-groove geometry, including geometric data.
	\label{fig:vgroovegeometry}}
\end{figure}

The findings from the previous section can be applied to other systems of different symmetry. 
For instance, a V-groove structure, i.e., a rod-shaped antenna with a notch on one side (see 
Fig.~\ref{fig:vgroovegeometry}) represents a system with a lower symmetry than that of a dimer
as it cannot be mirrored upside down. In a recent work we have found that this 
system exhibits strong second-harmonic generation efficiencies due to the possibility to engineer
the structure to exhibit double-resonant effects~\cite{hille2016second}.
In fact, the V-groove exhibits a symmetry corresponding to the point-group $C_2$. It is mapped 
onto itself by a rotation around the $y$-axis (which in two dimension is isomorphic to mirroring 
on the $yz$-plane). Since $C_2$ represents a subgroup of the $D_2$ group of the dimer, we can 
simply reuse the group-theoretical analysis developed for the dimer and adapt it to the V-groove
case. This means 
that the operations not belonging to $C_2$ have to be eliminated from the respective character table:
 \begin{center}
  \begin{tabular}{|>{\centering}m{1.5cm} | >{\centering}m{1.5cm}| >{\centering}m{1.5cm}| >{\centering}m{1.5cm} | > {\centering\arraybackslash}m{1.5cm}|}
\hline $ C_2$ & $I$ & $C_2(z)$ & $\sigma_{xz}$ & $\sigma_{yz}$ \\ \hline 
        $A$ & $1$ & - & - &  $1$ \\ \hline
      $B_1$ & $1$ & - & - & $-1$ \\ \hline 
  $B_2=B_1$ & $1$ & - & - & $-1$ \\ \hline
    $B_3=A$ & $1$ & - & - &  $1$ \\ \hline
\end{tabular}
\end{center}
Consequently, only two symmetries
as well as two classes of modes, hereafter called I' and II',
remain and we can identify the symmetries of the dimer geometry 
with those of the V-groove geometry as
\begin{align}
	B_{1}^{D_2} \equiv B^{{C_2}}, \\
	B_{2}^{D_2} \equiv B^{{C_2}}, \\
		A^{{D_2}} \equiv A^{{C_2}}, \\
	B_{3}^{D_2} \equiv A^{{C_2}}.
\end{align}
By carrying out the corresponding substitutions in the equations for the dimer, we readily obtain 
the symmetries for the first and second order fields for the V-groove.
Specifically, for a wave propagating in $y$-direction, with the $\vv{E}$-field polarized in $x$-direction, 
we find 
\begin{align}
                                  &\T{E^{k_y}_{1x,\mathrm{dimer}}}  =  A^{D_2} \oplus B_3^{D_2}  \notag \\
  \longrightarrow \quad &\T{E^{k_y}_{1x,\mathrm{v\mhyphen groove}}} =  A^{{C_2}}, \label{groove1lin}
 \end{align}
and, therefore, the second-order field is given by
\begin{align}
	\T{E^{k_y}_{2x,\mathrm{v\mhyphen groove}}}  &=  B^{{C_2}} \left[A^{{C_2}}\right]^2 \notag \\			
	&=  B^{{C_2}} \left(A^{{C_2}}\right)   \notag \\ 							
	&=  B^{{C_2}}. 
	\label{groove1nonlin}
\end{align}
This demonstrates that, for this direction of incidence, the nonlinearly excited modes are different 
from those that are excited
in the linear limit
.
\begin{figure}[ht]
	\centering
	\includegraphics[width=\linewidth]{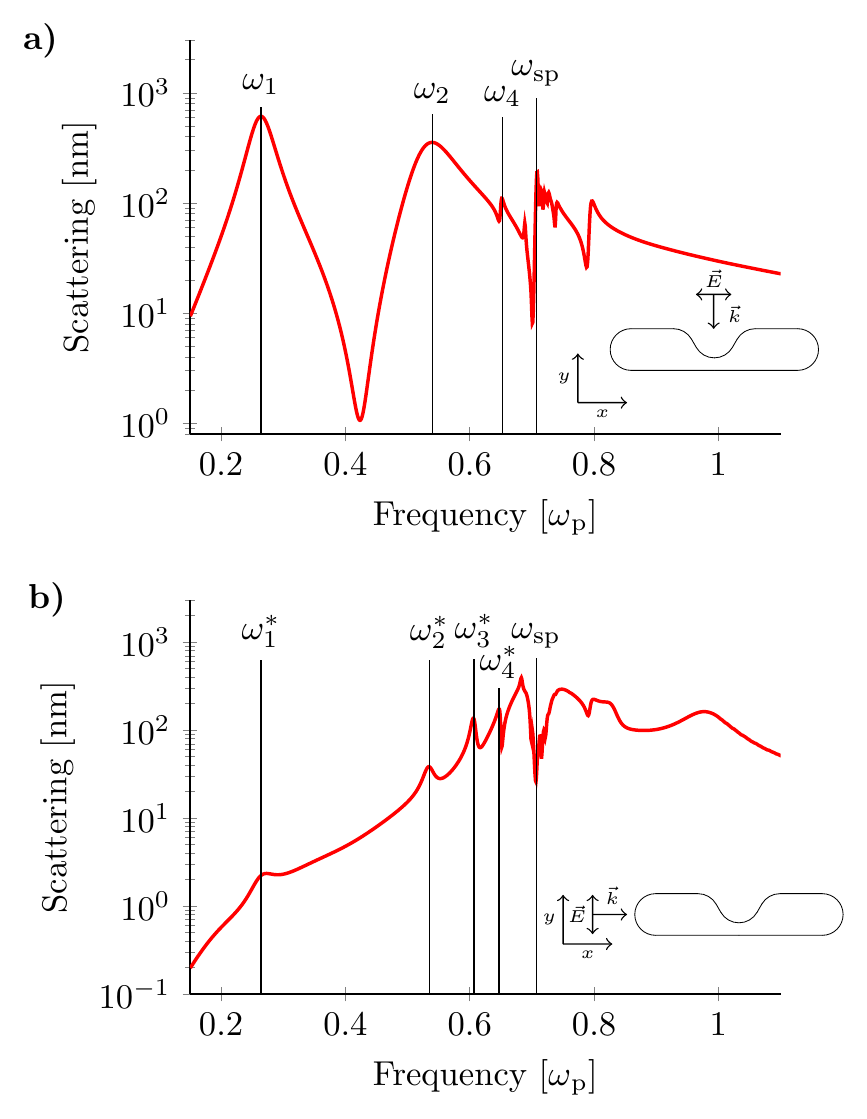}
	\caption{Linear scattering spectra for the V-groove structure. 
	         \captionbf{a)} Incidence along the short axis. 
					 \captionbf{b)} Incidence along the long axis. 
					 The resonances indicated by vertical lines will be 
           of importance in the discussion of SHG (see Fig.~\ref{fig:vgrooveshg}). 
					 The plots employ a logarithmic scale on the $y$-axis. 
					 \label{fig:vgroovelinear}}
\end{figure}
For pulses with direction of incidence along the $x$-axis, we find for the mode symmetries
\begin{align}
	                                 \T{E^{k_x}_{1x,\mathrm{dimer}}} &= B^{D_2}_1 \oplus B^{D_2}_3  \notag \\
  \longrightarrow \quad \T{E^{k_x}_{1x,\mathrm{v\mhyphen groove}}} &= B^{{C_2}} \oplus A^{{C_2}}, 
	\label{groove2lin}
\end{align}
so that
modes of both symmetry classes
can be excited. For the second-order fields, we obtain
\begin{align}
	\T{E^{k_x}_{2x},\mathrm{v\mhyphen groove}} &= B^{{C_2}} \left[ B^{{C_2}} \oplus A^{{C_2}} \right]^2  \notag \\
	&= B^{{C_2}} \left( A^{{C_2}} \oplus B^{{C_2}} \right)    \notag \\
                                             &=                  B^{{C_2}} \oplus A^{{C_2}}, 
	\label{groove2nonlin}
\end{align}
which demonstrates
that both classes of modes can also be excited at the second-order response
.

We are now in a position to computationally validate the symmetry relations, both
at the linear and second-order response level. In addition, we can classify the symmetries of the excited modes 
by close inspection of the corresponding linear and second-order scattering spectra. In other words: Contrary to
the standard approach where the symmetry of the excited modes allows for assessing of 
the second-order response, we have -- based on our group-theoretical analysis -- sufficient 
information to invert the procedure and determine the excited modes just by analyzing the 
linear and nonlinear spectra.

To demonstrate this, we again use a broad-band excitation pulse to compute the linear 
scattering spectra (cf. Fig.~\ref{fig:vgroovelinear}).
According to the symmetry analysis, for incidence with the $\vv{E}$-field polarized 
in $x$-direction and the $k$-vector pointing along the $y$-axis (\emph{short} axis), 
only one class of modes (class I') is excited. Thus, all modes in Fig.~\ref{fig:vgroovelinear}~\textbf{a)}, labeled $ \omega_1, \omega_2$ and 
$\omega_4$,  must belong to this class.
If the direction of incidence is rotated by 90 degrees (\emph{long} axis), both class I' and II' 
modes are excited. The latter spectrum is displayed in Fig.~\ref{fig:vgroovelinear}~\textbf{b)} and 
all resonances for the corresponding direction of incidence
$ \omega_1^*, \omega_2^*$, $\omega_3^*$ and $\omega_4^*$
are marked by an asterisk for later reference.
\begin{figure}[!b]
\includegraphics[width=\linewidth]{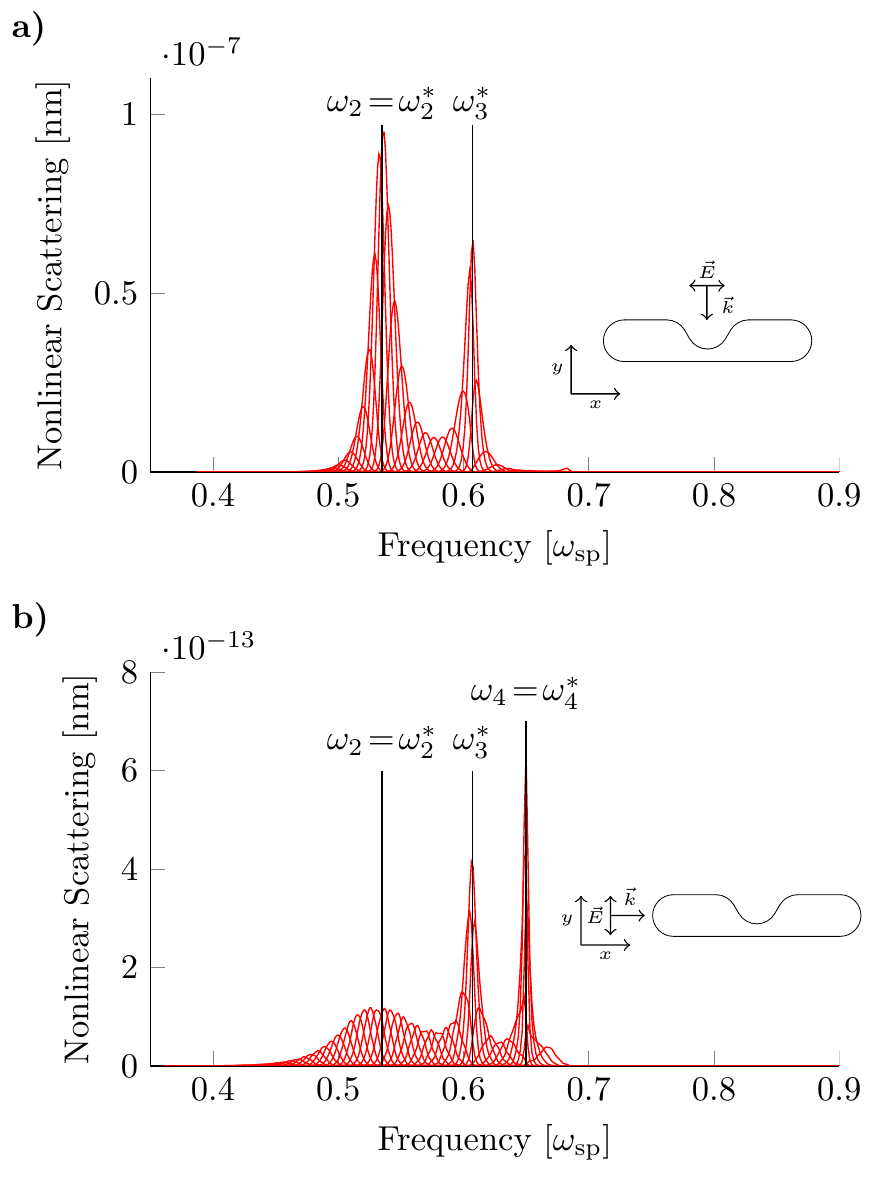}
\caption{Second-harmonic spectra from a double resonant V-groove.
         \captionbf{a)} Incidence along the short axis. A strong doubly-resonant behavior is found as 
				                the linear signal coincides with $\omega_1$, yielding a signal that is five 
						orders of magnitude larger than for the other angle of 
          incidence~\cite{hille2016second}.
          Resonances are found at the frequencies of $\omega_2=\omega_2^*$ and  that of $\omega_3^*$.
          \captionbf{b)} Incidence along the long axis. A double-resonant enhancement is not found as 
          the $\omega_1^*$ resonance on the linear level is weak for this direction of incidence.
          Resonances are found at the frequencies  $\omega_2=\omega_2^*$, of $\omega_3^*$ and  
          that of $\omega_4=\omega_4^*$. Note that the linear signal is only scanned across the 
          $\omega_1$ peak, from $0.25\,\omegap$ to $0.35\,\omegap$, where double-resonant behavior 
          is expected. This is not a frequency scan over the 
          whole spectrum.
          \label{fig:vgrooveshg}
}
\end{figure}
A close inspection of the spectra in Fig.~\ref{fig:vgroovelinear} reveals that for every frequency 
at which there is a resonance in \ref{fig:vgroovelinear}~\textbf{a)}, there is also a resonance in 
\ref{fig:vgroovelinear}~\textbf{b)}, but the reverse does not hold. This is fully consistent with 
the group-theoretical analysis which suggests that, for incidence along the long axis, both classes 
of modes are excited while for incidence along the short axis only one class is excited  
(see Eqs.~\eqref{groove1lin} and \eqref{groove2lin}). 
All peaks in the spectra of Fig.~\ref{fig:vgroovelinear}~\textbf{a)} must pertain to class I'.
Three of these peaks which will be important for our subsequent discussions are labeled by $\omega_1$, 
$\omega_2$ and $\omega_4$.
On the other hand, the peaks in Fig.~\ref{fig:vgroovelinear}~\textbf{b)} could correspond to either 
class I' or class II'
modes. We find that the peaks labeled by $\omega_1^*$, $\omega_2^*$ and $\omega_4^*$ occur at the 
same frequencies $\omega_1$, $\omega_2$ and $\omega_4$ as before.
Therefore, each of these peaks could have contributions of a class I' mode and of class II' mode.
Only for the peak labeled $\omega_3^*$,
there can be only contributions from class II', since no peak is found at this frequency when 
the system is excited along the short axis (Fig.~\ref{fig:vgroovelinear}~\textbf{b)}). By 
investigating the second-order response, the contributions to the peaks $\omega_1^*$, 
$\omega_2^*$ and $\omega_4^*$ can be identified.

According to Eq.~\eqref{eq:secondharmonicsymmetries}, for the second-order response, 
for incidence along the short axis, the nonlinearly excited modes are of class II', i.e., precisely those modes that are not excited by the 
fundamental signal. In turn, for 
incidence along the long axis, both on the fundamental (linear) and second-order (nonlinear) 
level, modes of class I' and II' will be excited. 

In Fig. \ref{fig:vgrooveshg}, we display the SHG spectra for the V-groove structure, just 
as it was done before for the dimer in Fig. \ref{fig:shg_spectra}. We notice first, that the frequencies $\omega_2^{*}$, $\omega_3^*$ and 
$\omega_4^{*}$
are such that they are approximately at twice the frequency of $\omega_1^{*}$. Thus, utilizing a fundamental pulse centered near 
$\omega_1^{*}$
that generates a second-harmonic signal at those peaks leads to a double-resonant excitation.
Note that since the fundamental signal is centered near $\omega_1^{*}$, the peak at 
$\omega_1^{*}$ itself is not included into the SHG-studies.
For incidence along the short axis, the fundamental signal at $\omega_1$ is very 
strong, which yields a strong double-resonance, while for incidence along the long axis,
the signal ($\omega_1^*$) is very small and, therefore, the double-resonance effect is 
weak (see Fig.~\ref{fig:vgrooveshg}). For incidence along the short axis, there is 
a strong SHG enhancement, yielding a signal which is approximately five orders of 
magnitude larger than the signal obtained for incidence along the long short axis. 

As pointed out above,
an analysis of the SHG spectra in
Fig. \ref{fig:vgrooveshg}
suffices to determine whether the peaks under investigation pertain to class I' or 
class II' (see Table \ref{symmetry} for a summary).
In fact, two very pronounced peaks at frequencies
$\omega_2^{*}$
and $\omega_3^*$ are found in Fig.~\ref{fig:vgrooveshg}~\textbf{a)}.
Above, we have already discussed that the class II' $\omega_3^*$ mode is expected 
in the SHG signal.
The fact that there is also a resonance at $\omega_2^{*}$ means that the $\omega_2^*$ 
peak in the linear spectrum must have a contribution of
a class II' mode which is degenerate in frequency with the $\omega_2$ class I' mode. 
On the other hand, the fact that there is no SHG signal at $\omega_4^*$ means that 
the $\omega_4^*$ 
has only a class I' contribution, just as $\omega_4$. 
The SHG signal for incidence along the long axis (Fig.~\ref{fig:vgrooveshg}~\textbf{b)}) 
exhibits peaks at all three frequencies $\omega_2^{*}$, $\omega_3^{*}$ and 
$\omega_4^{*}$. This confirms the group-theoretical prediction that, for this 
direction of incidence, modes from both symmetry classes are excited linearly and 
nonlinearly. Consequently, the mode at $\omega_3^*$, which is clearly of class II', 
is excited linearly and nonlinearly and so is the peak $\omega_4$ which, in turn,
we have identified as having contributions of class I'. 
Finally, the peak $\omega_2^{*}$ is a mixed result of the class I' and the class II' 
modes' response at this same frequency.

\begin{table}[ht]
\begin{center}
\begin{tabular}{c|c|c|c|c|cc}
 &$\omega_{2}$&$\omega_{2}^{*}$&$\omega_{3}^{*}$&$\omega_{4}$&$\omega_{4}^{*}$&\\
\hline
class I' & $\bullet$ &  $\bullet$ & &$\bullet$ &$\bullet$ &\\
class II' & &  $\bullet$ & $\bullet$ &   & &
\end{tabular}
\caption{
         Summary of the symmetry classes of the modes excited in the V-Groove 
				 geometry. The asterisk indicates that, in the linear spectrum, the 
				 spectral peak was found during excitation along the long axis of the 
				 structure. The peaks $\omega_{2}$ and $\omega_{2}^*$ are degenerate in 
				 frequency. While $\omega_{2}$ has only contributions of class I', 
         $\omega_{2}^*$ has contributions from both classes. The peak $\omega_{3}^{*}$, 
				 which was found only for one excitation direction, is of class II'. 
				 The peaks $\omega_{4}$ and $\omega_{4}^{*}$ come from the same mode, 
				 which is of class I'. The fundamental signal for this SHG-study was 
				 chosen near $\omega_{1}$ to exploit the double-resonant behavior of 
				 the structure, hence the peak at $\omega_{1}$ itself is excluded from 
				 the study. 
				 \label{symmetry}
				}
\end{center}
\end{table}

\section{Conclusions}
We have presented a numerical and analytical study of the plasmonic modes in different nanowire systems
and their excitation, on both the fundamental linear and second-order nonlinear level. 
A description based on the quasi-electrostatic approximation allows for an extremely useful mode 
classifications which, for cylindrical dimer systems and the ordinary Drude model have been carried 
out fully analytically. In combination with a group-theoretical analysis of the fundamental linear 
and second-order nonlinear response, we have established mode selection rules predicting the excitation 
of the linear and the nonlinear plasmonic modes for different directions of incidence. With the help 
of the analytical results, we have been able to interpret the resulting linear and nonlinear numerical 
spectra obtained using the DGTD method, although the computations were performed outside the quasi-electrostatic 
regime and using a nonlocal material model. 
In order to further demonstrate the utility of the mode classification and group-theoretical analyses,
we have derived the selection rules of a low-symmetry single V-groove system from those of the high-symmetry
dimer system. This also shows that our analyses can be readily applied to other geometries which 
have the same or a lower symmetry than a cylindrical dimer, i.e., a very wide range of possible nano-antenna
geometries.

Our analysis relies on the nonlocal, nonlinear and fully retarded hydrodynamic Drude 
model as described in Eqs. \eqref{HydroEqs}. In such a model the metal is described 
as a plasma with finite compressibility, in contrast zo a local description where 
the fluid is rigid. The hydrodynamic equations allow for the analytical inspections 
and numerical implementations presented above and provide a good description of the 
system. 
Further investigations will aim to improve such a description by considering further
effects such as the Landau damping, which were neglected in our present treatment.
Previous work~\cite{intravaia2015fluorescence} has indeed shown that the interplay 
between nonlocality and dissipation can substantially alter some features of the 
resonance spectrum, adding for example additional line-broadening. Modifications of 
the hydrodynamic model which partially take these features into account have already 
been presented in the literature
~\cite{toscano2015resonance,ciraci2016quantum,liebsch1993surface,toscano2015resonance,halevi1995hydrodynamic,mortensen2014generalized,raza2013nonlocal}. 
Further, by exploiting the connection to plasma physics, an alternative and highly
interesting procedure relies on the implementation of the fully nonlinear Boltzmann 
transport equation to describe the dynamics of the carrier (quantum) distribution 
in the metal
~\cite{manfredi2005model,bittencourt2013fundamentals}. 
The combination of such and other, similar approaches, providing a microscopically 
based description of the material properties, with an analysis of the geometry-induced 
system symmetries offers a powerful and constructive route for designing realistic 
and highly efficient nano-devices that are able to achieve strong SHG.

\section{Acknowledgments}
We acknowledge financial support by the Einstein Foundation Berlin (Project ActiPlAnt) and BMBF 
(Nano-Film, Project 13N14149). FI further 
acknowledges financial support from the DFG through the DIP program 
(Grant No. SCHM 1049/7-1). MM would like to thank Andreas Hille for inspiring discussions and 
for providing details on the V-groove structure.

%%%%%%%%%%%%%%%%%%%%%%
\newpage
\cleardoublepage
\setcounter{page}{1}

\newcounter{sfigure}
\setcounter{sfigure}{1}

\renewcommand{\theequation}{S\arabic{equation}}
\renewcommand{\thefigure}{S\arabic{sfigure}}

\newcommand{\partsymvec}{\left(\begin{array}{c} B_2 \\ B_3 \\ \end{array} \right)}

%\begingroup

\section*{\Large Supplemental Material}

In this supplementary document to the manuscript 
``\textit{Plasmonic modes in nanowire dimers: A comprehensive study based on the hydrodynamic
          Drude model including nonlocal and nonlinear effects}'', 
we present, in the first section, the symmetry considerations regarding the hydrodynamic 
equations coupled to the Maxwell equations in full detail. In the second section, we provide 
additional spectra and field distributions which supplement those displayed in the main text 
of the manuscript.
 
\section{Group-Theoretical Treatment of the Hydrodynamic Drude Model}
First, we recall the perturbative representation of the physical quantities in terms
of static fields, fields at the fundamental frequency, and fields corresponding to the
second-order response
\begin{align}
 \vv{E} &=  \vv{E}_0 + \vv{E}_1 + \vv{E}_2 + ... \\
 \vv{H} &=             \vv{H}_1 + \vv{H}_2 + ... \\
      n &=     n_0   +    n_1   +   n_2 +    ... \\
 \vv{v} &=             \vv{v}_1 + \vv{v}_2 + ... 
 \end{align}
They are inserted into the hydrodynamic equations (see Eqs. (17) and (18) of the main text) and 
Maxwell's equations. Then, the symmetries of all quantities are connected to each component by 
means of  these equations, starting with the symmetry of the incident field. The time-derivative 
does not alter the (spatial) symmetry. The point group corresponding to a cylindrical dimer structure 
is the dihedral group $D_2$. Accordingly, the symmetries of the nabla-operator are given by
\begin{align}
	\T{\nabla} = \begin{pmatrix} \T{\partial_x} \\ 
	                             \phantom{.} \\ 
	                             \T{\partial_y} \end{pmatrix} = 
							 \begin{pmatrix} B_2 \\ 
							                 B_3 \end{pmatrix}.
\end{align}
The first quantity which should be considered first within these equations is the equilibrium 
electron density $n_0$ which exhibits the same symmetry as the dimer itself
\begin{align}
	\T{n_0} = A.
\end{align} 
Without any further assupmtion, we can thus analyze the first-order continuity equation:
\begin{align}
			 \partial_t n_1      &= -\nabla \cdot (n_0 \vv{v_1})\\
	\Longrightarrow  \T{n_1} &= \partsymvec  A \T{\vv{v_1}}  \\
	\Longrightarrow  \T{n_1} &= B_2 \T{v_{1x}} \oplus B_3 \T{v_{1y}}, 
	\label{gl:eins}
\end{align}
which connects the linear term of to the density to the two components of the velocity-field.
The Euler equation yields
\begin{align}
	m_e n_0 \partial_t \vv{v}_1 &=  - m_e \gamma n_0 \vv{v}_1- \frac{5}{3} \kappa n_0^\frac{2}{3} \nabla n_1 -e n_0 \vv{E}_1 \\
\Longrightarrow \quad         & \T{v_{1x}} = B_2 \T{n_1} \oplus \T{E_{1x}}, \label{gl:zwei} \\
                              & \T{v_{1y}} = B_3 \T{n_1} \oplus \T{E_{1y}}. \label{gl:drei}
\end{align}
Inserting Eq.~\eqref{gl:eins} into Eqs.~\eqref{gl:zwei} and \eqref{gl:drei} leads to
\begin{align}
 \T{v_{1x}} &= B_2 B_2 \T{v_{1x}} \oplus B_2 B_3 \T{v_{1y}} \oplus \T{E_{1x}} \notag  \\
            &= B_1 \T{v_{1y}} \oplus \T{E_{1x}}, \label{gl:vier}\\
 \T{v_{1y}} &= B_1 \T{v_{1x}} \oplus \T{E_{1y}}, \label{gl:fuenf}
\end{align}
which allows to express the velocity components in terms of the electric fields
\begin{align}
  \T{v_{1x}} &= \T{E_{1x}} \oplus B_1  \T{E_{1y}}, \notag \\
  \T{v_{1y}} &= \T{E_{1y}} \oplus B_1  \T{E_{1x}}. 
	\label{gl:sieben}
\end{align}
These relations demonstate that the motion of the electrons given by $v_{1i}$ exhibits 
the same symmetry as $E_{1i}$ ($i=x,y$). This is reasonable since the electrons are 
accelerated by means of the electric field. This is easily fulfilled if the second
field component $E_{1j}$ ($j=y,x$) is zero. 
In the case of the dimer modes, $E_{1j}\neq 0$, but then $\T{E_{1j}}$ is not independent 
of $\T{E_{1i}}$. The two are connected through the electrostatic potential. Since the 
gradient operator is given by  
$\T{\nabla} = \left(\,  B_2 \,\,\,  B_3 \, \right)^T$, the relation 
\begin{align}
	\T{E_{1x}} = B_2 \T{V} = B_1 B_3 \T{V} =  B_1 \T{E_{1y}} 
	\label{gl:sternchen}
\end{align}
holds, in agreement with Eq.~\eqref{gl:sieben}. 

Another consistency check of the procedure can be realized by combining Eq.~\eqref{gl:sieben} 
with the continuity equation, Eq.~\eqref{gl:eins}. We then find
\begin{align}
	\T{n_1}= B_2 \T{E_{1x}} \oplus B_3\T{E_{1y}}. 
	\label{gl:acht}
\end{align}
Eq.~\eqref{gl:acht} shows that the first order density has the same
symmetry as the electrostatic potential 
\begin{align}
	\T{n_1} = \T{V},
\end{align}
which is certainly fulfilled, because the excess charges give rise to the potentials and, 
therefore, must exhibit the same symmetries. 
\\
Before we move on to the second order equations, we need to determine the symmetry of the 
magnetic field (for the polarization employed here, there is only the $H_z$-component),
which is a necessary ingredient for the second order Euler equation. 
From Maxwell's curl equations, we can straightforwardly extracted
\begin{align}
				 \partial_t \vv{H_1} &= -\nabla \times \vv{E}_1 \\
	\Longrightarrow \T{H_{1z}} &= B_3 \T{E_{1x}} \oplus B_2\T{E_{1y}}.
\end{align}

\begin{widetext}

The second-order continuity equation reads
\begin{align}
								 \partial_t n_2 &= - \nabla(n_1 \vv{v}_1 + n_0 \vv{v}_2) \\
  \Longrightarrow \quad \T{n_2} &= \T{n_1}             \left[ B_2 \T{v_{1x}} \oplus B_3 \T{v_{1y}} \right]  \notag \\
                                & \phantom{=} \oplus A \left[ B_2 \T{v_{2x}} \oplus B_3 \T{v_{2y}} \right] \\
  \stackrel{\ref{gl:acht},\ref{gl:sieben}}{\Longrightarrow} \quad \T{n_2}& =\left[ B_2 \T{E_{1x}} \oplus B_3\T{E_{1y}}\right]^2 \oplus B_2 \T{v_{2x}} \oplus B_3 \T{v_{2y}} \\
  \stackrel{\ref{gl:sternchen}}{\Longrightarrow}\quad \T{n_2} &=\left[ \T{E_{1x} }\right]^2 \oplus B_2 \T{v_{2x}} \oplus B_3 \T{v_{2y}} \label{gl:neun}
\end{align}
The second-order Euler equation is given by 
\begin{align}
  &m_e (n_0 \partial_t \vv{v}_2) + m_0n_1\partial_t\vv{v_1} + m_e n_0 (\vv{v}_1\cdot \nabla) \vv{v}_1  \notag \\
  = &-m_e \gamma (n_1\vv{v_1}+n_0\vv{v_2} )  \notag\\
  &-\frac{5}{9} \kappa n_0^{-1/3} \nabla n_1^2 - \frac{5}{3}\kappa n_0^{2/3}\nabla n_2 \notag \\
  &-e (n_1 \vv{E}_1 + n_0 \vv{E}_2) \notag \\
  &-e n_0 \vv{v}_1 \times \vv{H}_1 \\
  \stackrel{x\mathrm{-component}}{\Longrightarrow} \quad 
  &\T{v_{2x}} \oplus \T{n_1}\T{v_{1x}} + \left[ B_2 \T{v_{1x}} \oplus B_3 \T{v_{1y}} \right] \T{v_{1x}} \notag \\
  = &\T{n_1}\T{v_{1x}}  \oplus \T{v_{2x}} \oplus B_2 \left[\T{n_1} \right]^2 \oplus B_2 \T{n_2} \notag \\
    &\oplus  \T{n_1} \T{E_{1x}} \oplus \T{E_{2x}} \oplus \T{v_{1y}} \T{H_{1z}} \\
    \stackrel{\ref{gl:sternchen}\ref{gl:sieben}}{\Longrightarrow} \quad
    B_2 \left[ \T{E_{1x}} \right]^2 &= B_2 \left[ \T{E_{1x}} \right]^2 \oplus \T{v_{2x}}  \oplus B_2 \left[ \T{E_{1x}} \right]^2 \oplus B_2 \T{n_2} \notag \\
    &\phantom{=}  \oplus B_2 \left[ \T{E_{1x}} \right]^2 \oplus \T{E_{2x}} \oplus B_1 \T{E_{1x}} \T{H_{1z}} \notag \\
    &=  \T{v_{2x}} \oplus  B_2 \T{n_2}  \oplus \T{E_{2x}}.
\end{align}
\end{widetext}

This results in an expression for  $\T{v_{2x}}$ and by repeating the same procedure for the $y$-component, 
we arrive at
\begin{align}
  \T{v_{2x}} =  B_2 \left[ \T{E_{1x}} \right]^2 \oplus B_2 \T{n_2}  \oplus \T{E_{2x}} \\
  \T{v_{2y}} =  B_3 \left[ \T{E_{1x}} \right]^2 \oplus B_3 \T{n_2}  \oplus \T{E_{2y}}.
\end{align}
Inserting both into Eq.~\eqref{gl:neun} yields
\begin{align}
	\T{n_2} = \left[ \T{E_{1x}} \right]^2 \oplus B_2 \T{E_{2x}},
\end{align}
which establishes a connection between the second-order density, second-order electric field and the 
first-order electric field. Further, the second-order density and the second-order electric field are 
connected via  Gau\ss' law
\begin{align}
	B_2 \T{E_{2x}} \oplus B_3 B_1 \T{E_{2x}} =\T{n_2},
\end{align}
where we made use of the fact that the second-order electric field components are connected in the 
same way as those of first-order because within the curl-equations, the different orders decouple.
As a result, we can write down the relation between the first-order and the second-order electric 
field
\begin{align}
	\T{E_{2x}} =  B_2 \left[\T{E_{1x}}\right]^2.
\end{align}

\newpage

\section{Absorption spectra and field images}

Within the main text, we have discussed the effects of nonlocality on the scattering spectra. Here, 
we display the corresponding absorption spectra for the same setup. 

\begin{figure*}
	\includegraphics[width=0.75\textwidth]{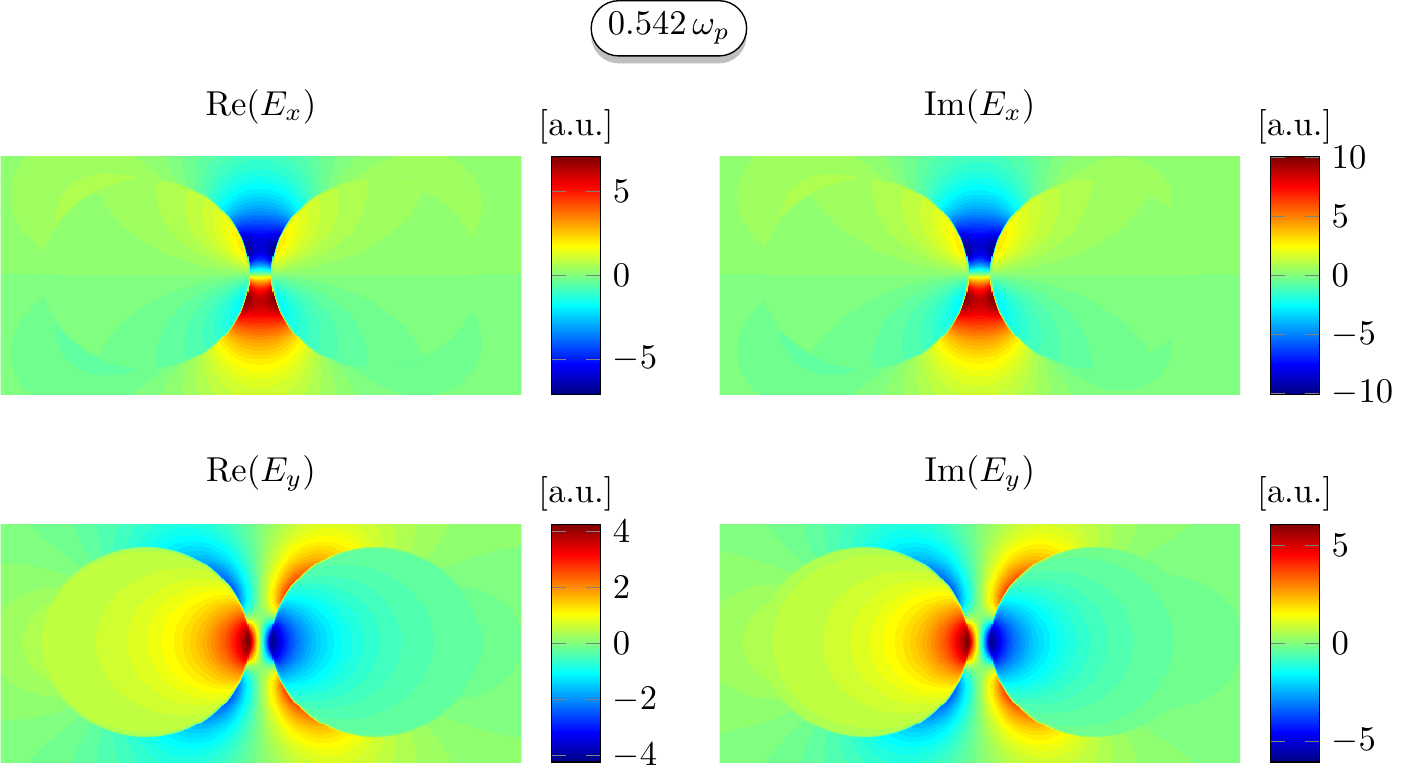}
	\caption{Field distributions (real and imaginary part) for the $m=1$ mode at $0.542 \omega_\mathrm{p}$. 
	         Both modes are of class I.
	         \label{fig:90deg_mode1_low}}
 \stepcounter{sfigure}{1}
\end{figure*}

\begin{figure*}
	\includegraphics[width=0.75\textwidth]{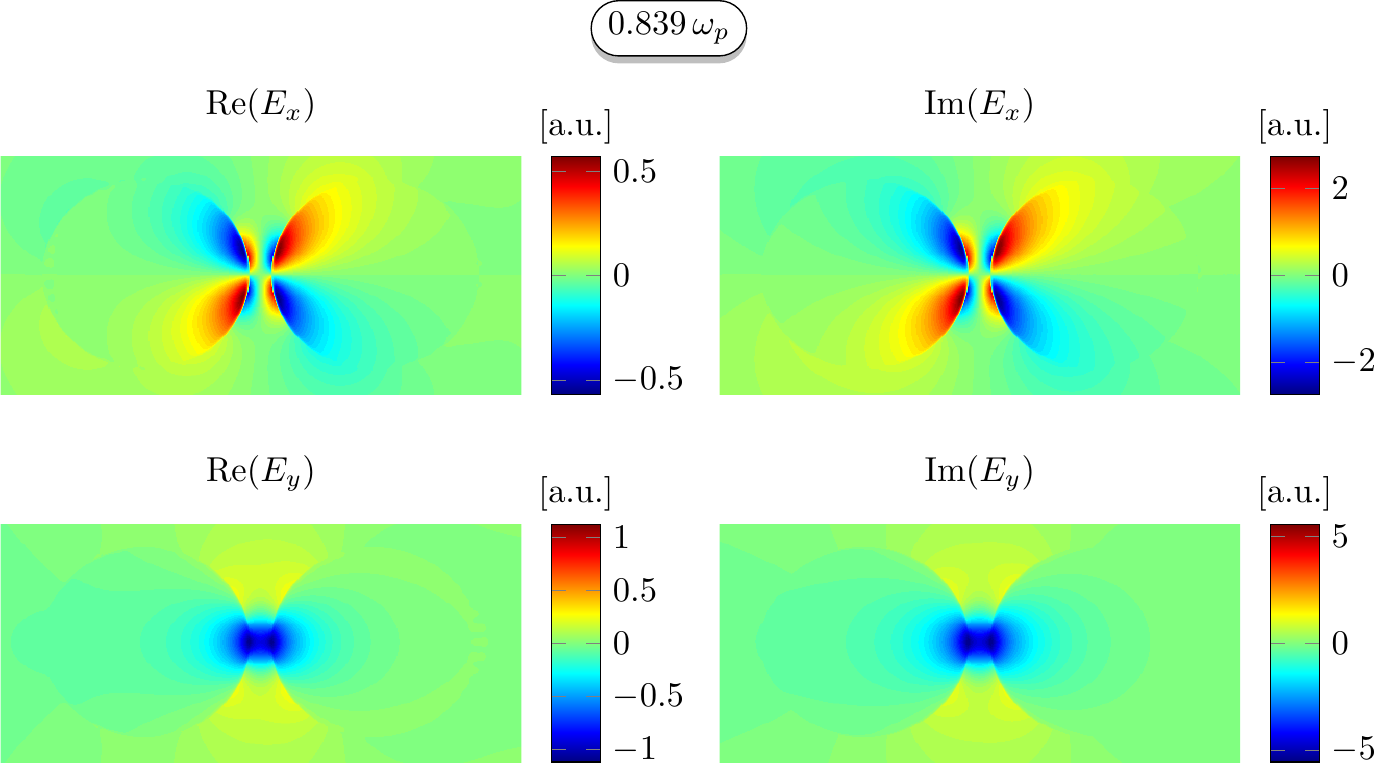}
	\caption{Field distributions (real and imaginary part) for the $m=1$ mode at $0.839 \omega_\mathrm{p}$. 
					 Both distributions are of class II.
					 \label{fig:90deg_mode1_high}}
 \stepcounter{sfigure}{1}
\end{figure*}

\begin{figure*}
	\includegraphics[width=0.75\textwidth]{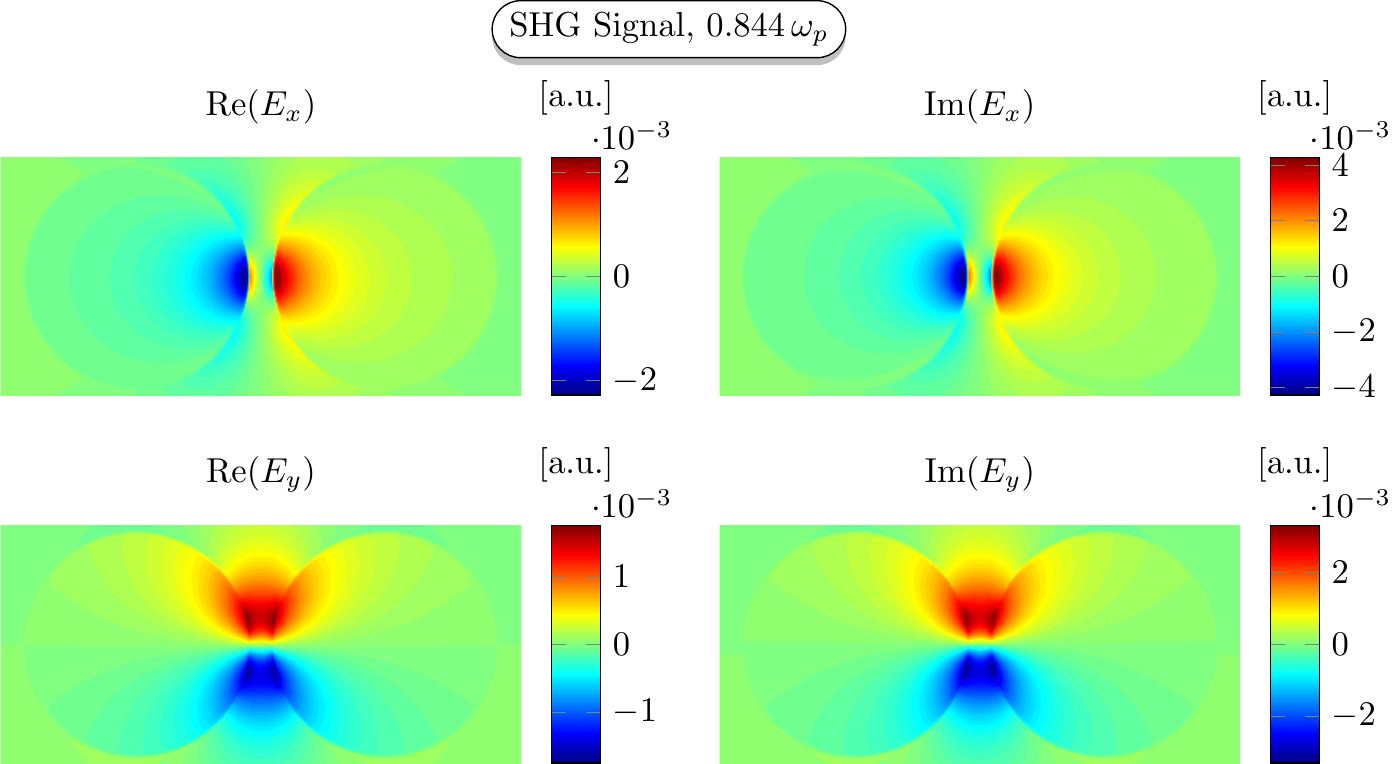}
	\caption{Field distributions (real and imaginary part) for the $m=1$ mode at $0.844 \omega_\mathrm{p}$, 
	         excited through second-order response. Both distributions are of class IV.
	         \label{fig:SHG_0deg_mode1_high}}
 \stepcounter{sfigure}{1}
\end{figure*}

\begin{figure*}
	\includegraphics[width=0.75\textwidth]{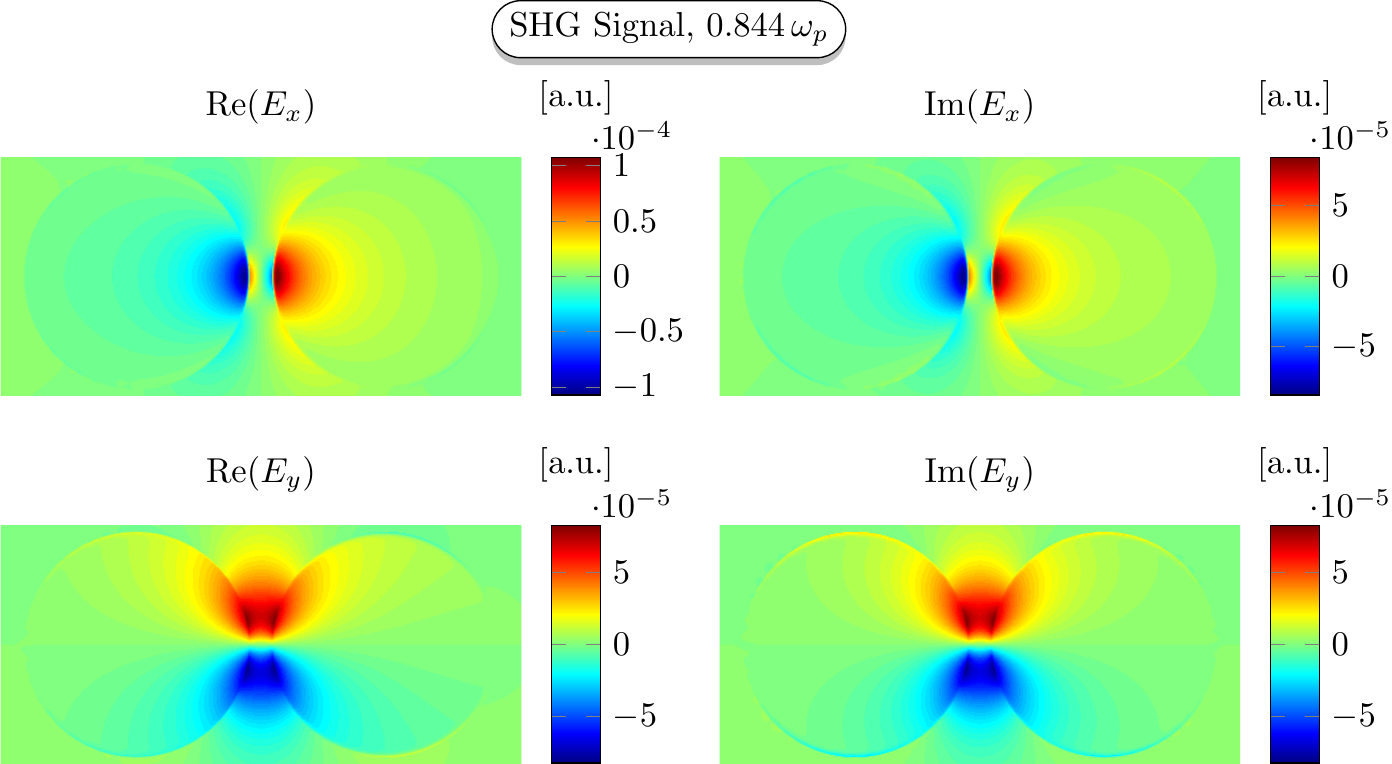}
	\caption{Field distributions (real and imaginary part) for the $m=1$ mode at $0.844 \omega_\mathrm{p}$, 
	        excited through second-order response. Both are of class IV.
	        \label{fig:SHG_90deg_mode1_high}}
 \stepcounter{sfigure}{1}
\end{figure*}

Furthermore, in the main text, we have presented the real parts of the electric field distributions, 
which have allowed us to identify the modes. Here, we also display the imaginary parts of the fields 
(all of them for nonlocal computations). In all cases discussed in the main text, the imaginary part 
pertains to the same symmetry class as the real part (Figs.~\ref{fig:0deg_mode1}, \ref{fig:90deg_mode1_low},
\ref{fig:90deg_mode1_high}, \ref{fig:SHG_0deg_mode1_high}, \ref{fig:SHG_90deg_mode1_high}), whereas, 
in the case of Fig.~\ref{fig:0deg_mode1and2mix}, we find that the real part is of class III while the 
imaginary part is of class I.

%\begin{widetext}

%\subsection{Fundamental signal}

\begin{figure*}
	\includegraphics[width=0.75\textwidth]{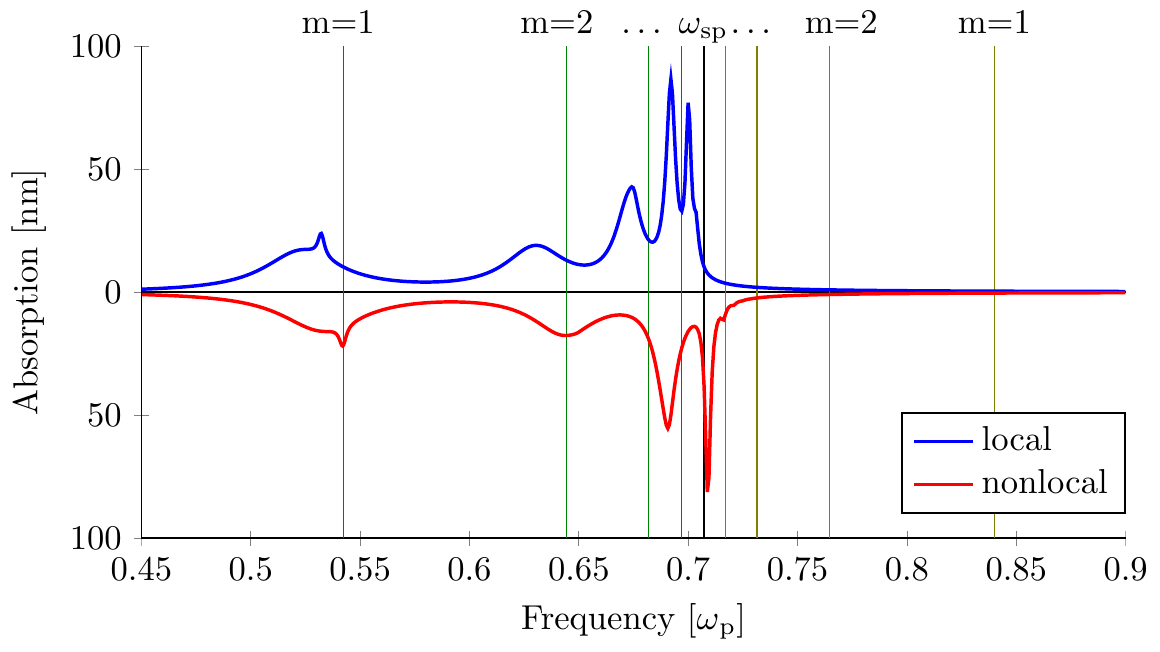}
	\caption{Absorption spectra for a cylindrical nano-wire dimer, for incidence perpendicular to the 
	         long dimer axis. For $m=1$, near $.54\omega_\mathrm{p}$, we observe a sharp peak within 
					 a broad peak. They belong to classes I (sharp peak) and III (broad peak), the degeneracy 
					 which is found in the quasi-electrostatic case is lifted in our numerically exact 
					 time-domain computations of the Maxwell equations.\label{fig:abso_0deg}}
					 \stepcounter{sfigure}{1}
\end{figure*}

\begin{figure*}
	\includegraphics[width=0.75\textwidth]{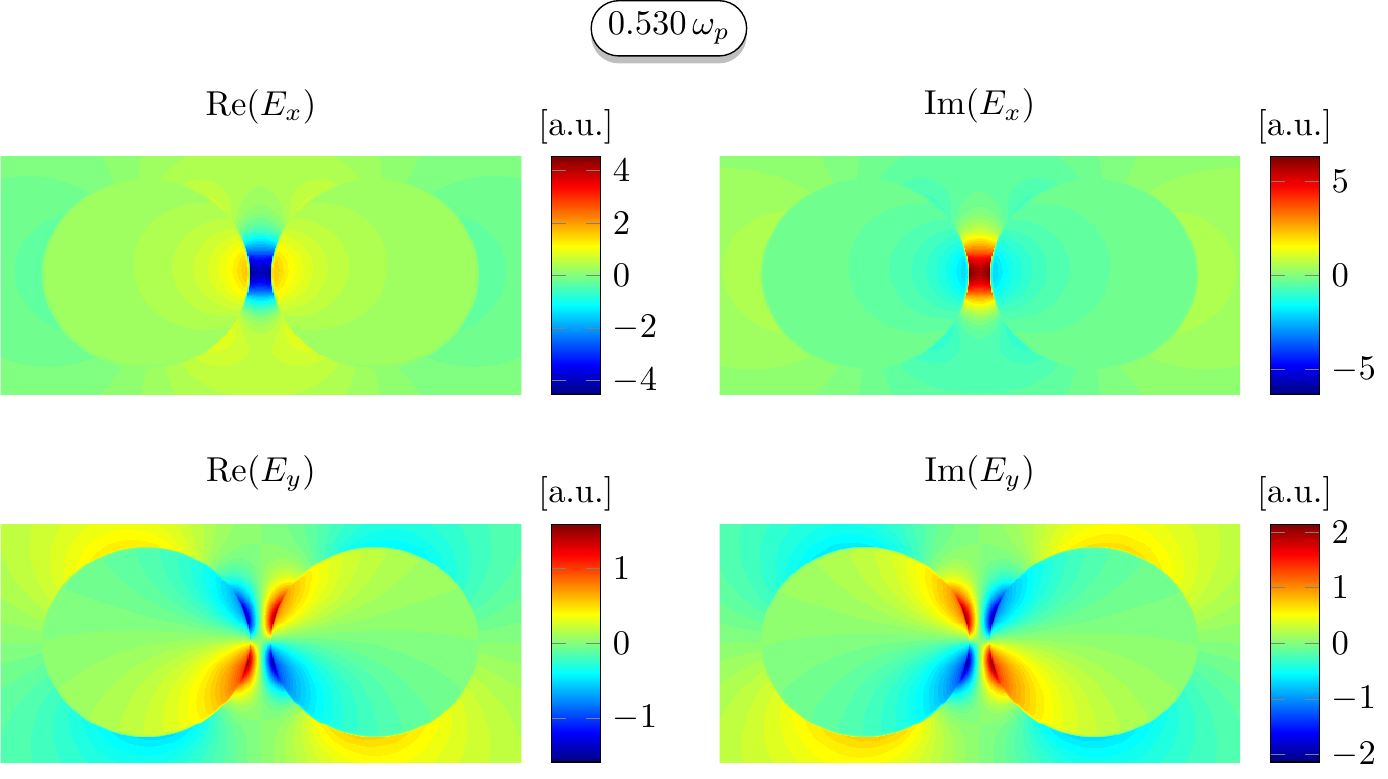}
	\caption{Field distributions (real and imaginary part) at the frequency corresponding to the 
	         maximum of the broad peak for $m=1$ (cf. Fig.~\ref{fig:abso_0deg}). Both, the real and 
					 the imaginary part, belong to class III. \label{fig:0deg_mode1} }
 \stepcounter{sfigure}{1}
\end{figure*}

\begin{figure*}
	\includegraphics[width=0.75\textwidth]{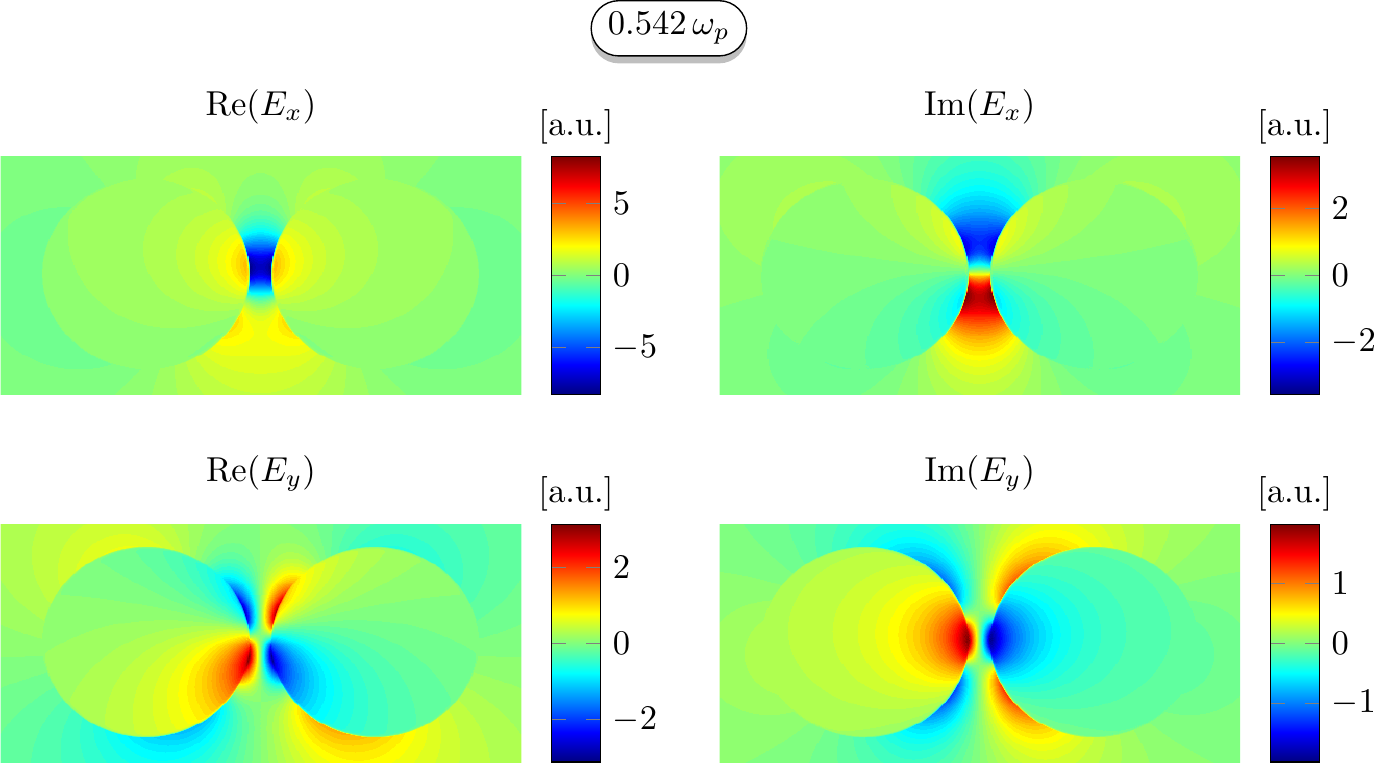}
	\caption{Field distributions (real and imaginary part) at the frequency corresponding to the 
	         maximum of the sharp peak for $m=1$ (cf. Fig.~\ref{fig:abso_0deg}). The real part is 
					a (slightly distorted) class III mode, while the imaginary part is of class I.
					At this frequency, the contributions of both modes overlap. \label{fig:0deg_mode1and2mix}}
 \stepcounter{sfigure}{1}
\end{figure*}

\begin{figure*}
	\includegraphics[width=0.75\textwidth]{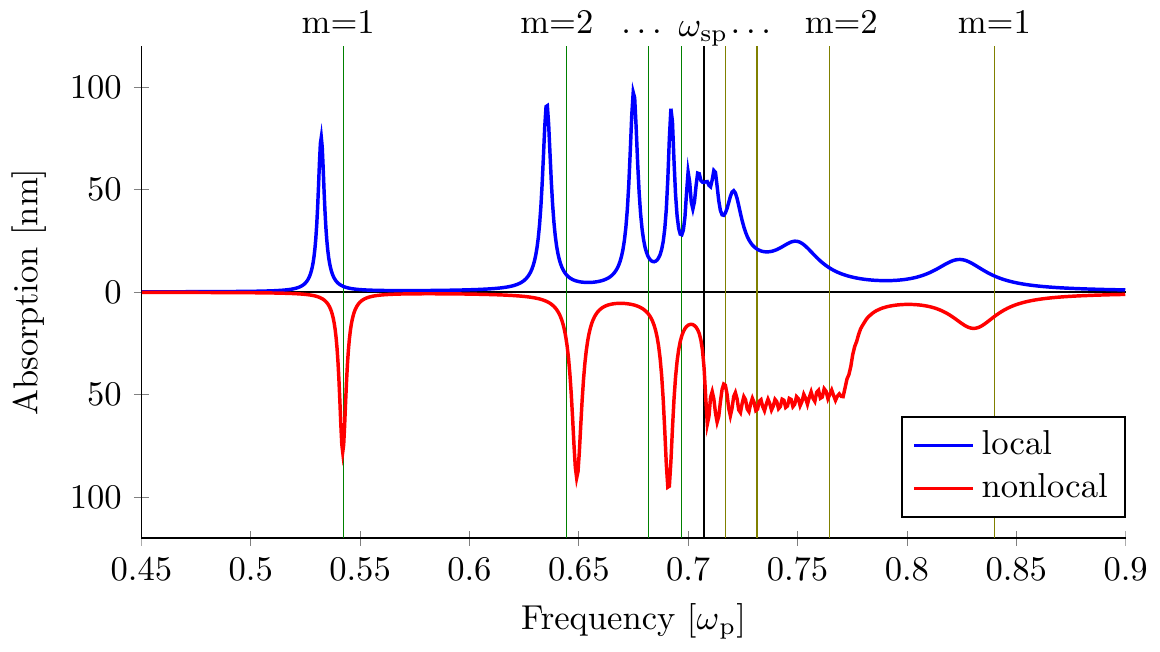}
	\caption{Absorption spectra for a cylindrical nano-wire dimer, for incidence along the long 
	         dimer axis. As for the scattering spectra, the nonlocality has a strong effect for 
					 the higher-order modes just above the surface plasmon frequency $\omega_{\mathrm{sp}}$.
					 \label{fig:abso_90deg}}
  \stepcounter{sfigure}{1}
 \end{figure*}

\begin{figure*}
	\includegraphics[width=0.75\textwidth]{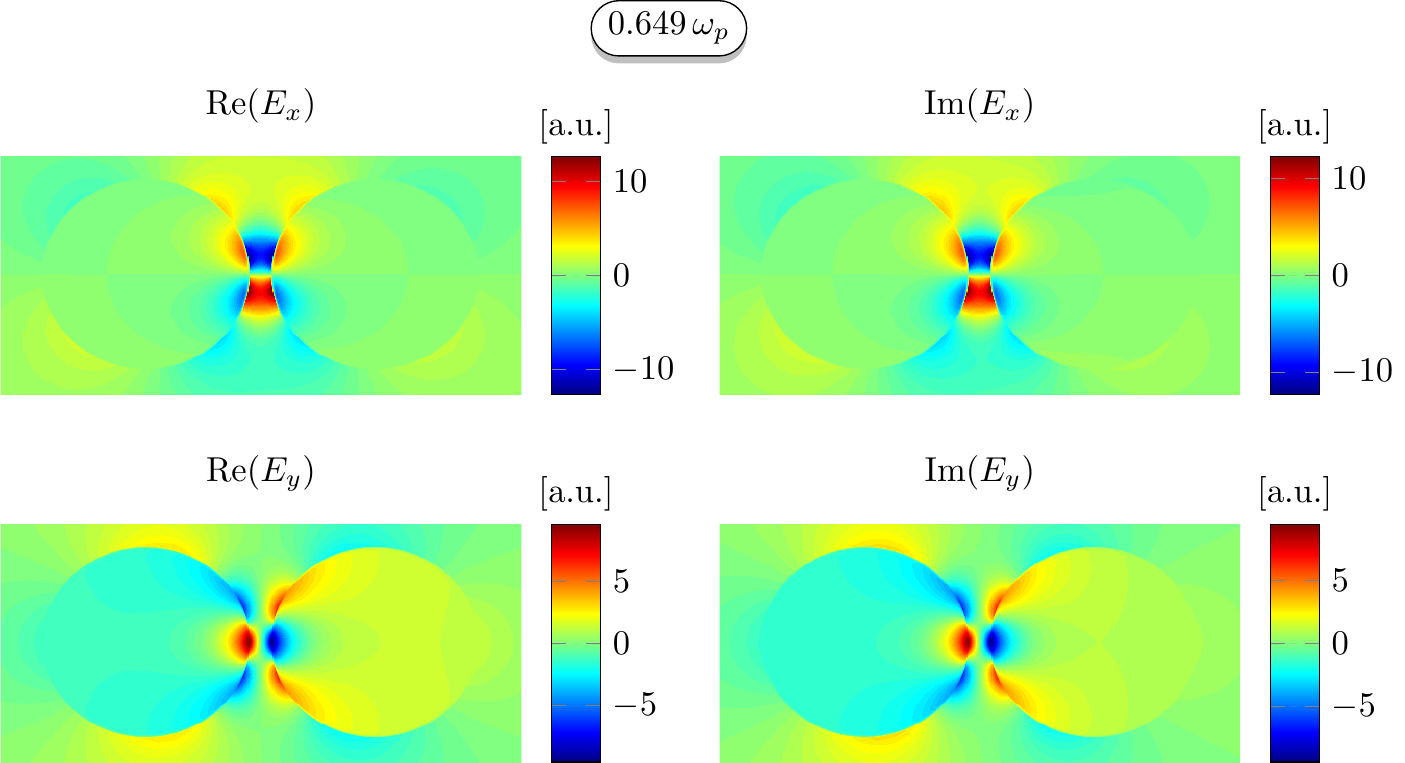}
	\caption{Field distributions (real and imaginary part) for the mode at $0.649 \omega_\mathrm{p}$. 
	         They are of class I, with $m=2$.
	         \label{fig:90deg_mode2_low}}
 \stepcounter{sfigure}{1}
\end{figure*}

\begin{figure*}
	\includegraphics[width=0.75\textwidth]{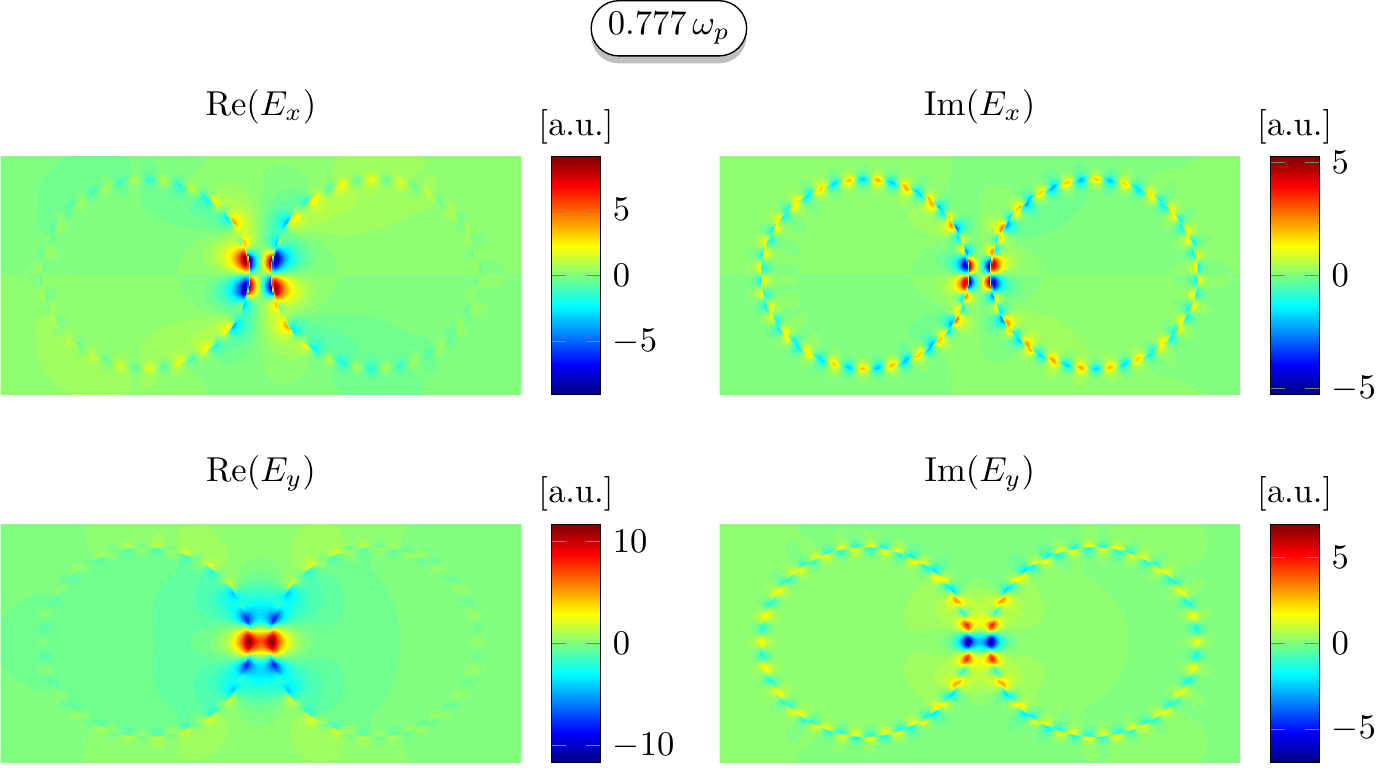}
	\caption{Field distributions (real and imaginary part) for the mode at $0.777 \omega_\mathrm{p}$. 
	         In the gap-region between the cylinders they exhibit certain characteristics of a 
					 class II mode with $m=2$, but the mode is not clearly separated and higher-order
					 modes contribute. The numerous minima and maxima on the surface of the cylinders are 
					 characteristic for large $m$.
					 \label{fig:90deg_middle}}
 \stepcounter{sfigure}{1}
\end{figure*}

\begin{itemize}
\item{\bf Incidence perpendicular to the long dimer axis:}
See Figs. \ref{fig:abso_0deg}, \ref{fig:0deg_mode1} and \ref{fig:0deg_mode1and2mix}.
\item {\bf Incidence parallel to the long dimer axis:}
See Figs. \ref{fig:abso_90deg}, \ref{fig:90deg_mode1_low}, \ref{fig:90deg_mode2_low}, \ref{fig:90deg_middle} and \ref{fig:90deg_mode1_high}.
\item {\bf Second-harmonic generation: Incidence perpendicular to the long dimer axis:}
See Fig. \ref{fig:SHG_0deg_mode1_high}.
\item {\bf Second harmonic generation: Incidence parallel to the long dimer axis:} See Fig. \ref{fig:SHG_90deg_mode1_high}.
\end{itemize}

%
%\clearpage
%\newpage

%\subsubsection{\bf Incidence parallel to the long dimer axis}

% 
 
%\clearpage
%\newpage
 
%\subsubsection{\bf Second-harmonic generation: Incidence perpendicular to the long dimer axis}

%\clearpage
%\newpage

%\subsubsection{\bf Second harmonic generation: Incidence parallel to the long dimer axis}

%\end{widetext}
%\endgroup

\end{document}